\documentclass[12pt,a4paper]{article} %\input epsf.tex
\pdfoutput=1

\usepackage[utf8]{inputenc}
\usepackage[T1]{fontenc}
\usepackage{lmodern, amsmath, amssymb, slashed, graphicx, adjustbox}
\usepackage[dvipsnames]{xcolor}
\usepackage[numbers, elide, compress]{natbib}
\usepackage[colorlinks=true, citecolor={blue}, urlcolor={blue}, linktocpage=true, linkcolor=blue, breaklinks=true, pdfpagelabels=false]{hyperref}

\usepackage{attachfile}

\usepackage[font={small}]{caption}% Writes captions of figures in small font

\graphicspath{{figs/}}

\makeatletter\g@addto@macro\bfseries{\boldmath}\makeatother

\def\figureautorefname~#1\null{Fig.\,#1\null}

\def\equationautorefname~#1\null{Eq.\,(#1)\null}

\normalsize
\numberwithin{equation}{section}

\newcommand{\inab}{\,{\rm ab}^{-1}}
\newcommand{\infb}{\,{\rm fb}^{-1}}

\newcommand{\eehz}{e^+e^- \to hZ}
\newcommand{\eevvh}{e^+e^- \to \nu \bar{\nu} h}
\newcommand{\eeww}{e^+e^- \to WW}

\newcommand{\vvh}{\nu \bar{\nu} h}

\newcommand{\eeeeh}{e^+e^- \to e^+ e^- h}

\newcommand{\eezhh}{e^+e^- \to Zhh}
\newcommand{\eevvhh}{e^+e^- \to \nu \bar{\nu} hh}

\newcommand{\vvhh}{\nu \bar{\nu} hh}

\newcommand{\eetth}{e^+e^- \to t \bar{t} h}

\newcommand{\dkl}{\delta \kappa_\lambda}
\newcommand{\kl}{\kappa_\lambda}

\newcommand*{\dzh}{\delta\hspace{-1pt}Z_H}

\newcommand{\bpm}{\begin{pmatrix}}
\newcommand{\epm}{\end{pmatrix}}

\begin{document}

\begin{flushright}
DESY 17-131 \\
FERMILAB-PUB-17-462-T
\end{flushright}

\vspace*{1cm}

\begin{center}

{\Large\bf
A global view on the Higgs self-coupling\\ at lepton colliders
\par}

\vspace{9mm}

{\bf Stefano~Di~Vita,$^{a,b}$ Gauthier~Durieux,$^{b}$ Christophe~Grojean,$^{b,c}$ \footnote{On leave from Instituci\'o Catalana de Recerca i Estudis Avan\c cats, 08010 Barcelona, Spain} Jiayin~Gu,$^{b,d}$ Zhen~Liu,$^{e}$ Giuliano~Panico,$^{f}$ Marc~Riembau,$^{b,f}$ Thibaud~Vantalon\,$^{b,f}$ }\\ [4mm]
{\small\it
$^a$ INFN Sezione di Milano, via Celoria 16, I-20133 Milano, Italy \\[2mm]
$^b$ DESY, Notkestra{\ss}e 85, D-22607 Hamburg, Germany \\[2mm]
$^c$ Institut f\"ur Physik, Humboldt-Universit\"at zu Berlin, D-12489 Berlin, Germany  \\[2mm]
$^d$ Center for Future High Energy Physics, Institute of High Energy Physics, \\ Chinese Academy of Sciences, Beijing 100049, China \\[2mm]
$^e$  Theoretical Physics Department, Fermi National Accelerator Laboratory, Batavia, IL, 60510 \\[2mm]
$^f$   IFAE and BIST, Universitat Aut\`onoma de Barcelona, E-08193~Bellaterra,~Barcelona,~Spain
\par}
\vspace{.5cm}
\centerline{\tt \small stefano.divita@mi.infn.it, gauthier.durieux@desy.de, christophe.grojean@desy.de,}
\centerline{\tt \small jiayin.gu@desy.de, zliu2@fnal.gov, gpanico@ifae.es, marc.riembau@desy.de,}
\centerline{\tt \small tvantalon@ifae.es}

\end{center}

\begin{abstract}
We perform a global effective-field-theory analysis to assess the precision on the determination of the Higgs trilinear self-coupling
at future lepton colliders. Two main scenarios are considered, depending on whether the center-of-mass energy
of the colliders is sufficient or not to access Higgs pair production processes. 
Low-energy machines allow for $\sim 40\%$ precision
on the extraction of the Higgs trilinear coupling through the exploitation of next-to-leading-order effects in single Higgs
measurements, provided that runs at both $240/250\,$GeV and $350$\,GeV are available with luminosities in the few attobarns range. A global fit, including possible deviations in other SM couplings,
is essential in this case to obtain a robust determination of the Higgs self-coupling.
High-energy machines can easily achieve a $\sim 20\%$ precision through Higgs pair production processes. In this case,
the impact of additional coupling modifications is milder, although not completely negligible.
\end{abstract}

\newpage
{\small 
\tableofcontents}

\setcounter{footnote}{0}

%%%%%%%%%%%%%%%%%%%%%%%%%%%%%%%%%%%%%%%%%%%%%%%%%%%%%%%%%%%%%%%%%%
%%%%%%%%%%%%%%%%%%%%%%%%%%%%%%%%%%%%%%%%%%%%%%%%%%%%%%%%%%%%%%%%%%
\pagebreak
\section{Introduction}

So far, the LHC provided us with a good deal of information about the Higgs boson. The determination of its linear couplings
to several Standard Model (SM) particles is nowadays approaching, and in some cases surpassing, the $10\%$ precision,
allowing for powerful probes of a broad class of natural beyond-the-SM (BSM) theories. On the contrary, the prospects for
measuring the Higgs self interactions, namely its trilinear and quadrilinear self-couplings, are much less promising.
At present, the trilinear Higgs coupling is loosely constrained at the $\mathcal{O}(10)$ level,
and the high-luminosity LHC (HL-LHC) program
could only test it with an ${\cal O}(1)$ accuracy (see for instance the experimental projections in Refs.~\cite{ATL-PHYS-PUB-2014-019,ATL-PHYS-PUB-2017-001}). The prospects for extracting the quadrilinear Higgs self-coupling are even less promising. 

From a theoretical point of view, on the other hand, the determination of the Higgs self-interactions is of primary importance.
They characterize
the Higgs potential, whose structure could shed some light on the naturalness problem.
Moreover, they control the properties of the electroweak phase transition, determining its possible relevance for baryogenesis.
Sizable deviations in the Higgs self-couplings are expected in several BSM scenarios, including for instance Higgs portal
models or theories with Higgs compositeness. All these considerations motivate the effort spent investigating the achievable
precision on the Higgs self-interactions at future collider experiments.

Projections for high-energy hadron machines ($100\,$TeV $pp$ colliders in particular) are already available in the
literature~\cite{Contino:2016spe}. They show that a very good precision on the determination of the trilinear Higgs coupling,
of the order of $5\%$, is possible. High-energy hadron machines, however, might only be constructed in a distant future and
could be preceded by lower-energy lepton colliders. It is thus worth studying the impact of future
lepton machines on the determination of the Higgs potential. In this work, we perform such an analysis, providing an
assessment of the achievable precision on the determination of the Higgs trilinear self-coupling.

We consider a comprehensive set of benchmark scenarios including low-energy lepton machines (such as FCC-ee and CEPC)
as well as machines that can also run at higher energies (ILC and CLIC). We will show that low-energy colliders, although not able to access directly
the Higgs trilinear coupling in Higgs pair production processes, can still probe it by exploiting loop corrections to single
Higgs channels that can be measured to a very high precision.
This approach, pioneered in Ref.~\cite{McCullough:2013rea}, allows for a good determination of the Higgs trilinear
interaction, which can easily surpass the HL-LHC one.
In performing this analysis, however, one must cope with the fact that different new-physics effects may affect simultaneously
the single Higgs cross sections, see Ref.~\cite{Durieux:2017rsg} as well as Refs.~\cite{Craig:2014una,Henning:2014gca, Ellis:2015sca, Ge:2016zro, deBlas:2016ojx, Ellis:2017kfi, Khanpour:2017cfq, Barklow:2017suo,Barklow:2017awn}.
In such a situation, a robust determination of the Higgs self coupling can only be obtained through a global fit that takes into
account possible deviations in other SM couplings. We will show that, within the SM effective field theory (EFT) framework with
a mild set of assumptions, the relevant operators correcting single Higgs production can be constrained provided enough
channels are taken into account.
In this way, a consistent determination of the Higgs self-coupling is possible even without direct access to Higgs pair production.

High-energy machines, on the other hand, are able to directly probe the trilinear coupling via Higgs pair production,
through $Zhh$ associated production and $WW$-fusion. We will see that these two channels provide complementary information
about the Higgs self interaction, being more sensitive to positive and negative deviations from the SM value respectively.
We will also show, as anticipated in Ref.~\cite{Contino:2013gna}, that a differential analysis of the $WW$-fusion channel, taking into account the Higgs-pair invariant mass
distribution, can be useful to constrain sizable positive deviations in the Higgs trilinear coupling that are hard to probe with an inclusive study.

The paper is organized as follows. In \autoref{sec:single}, we discuss the indirect trilinear Higgs coupling determination through single-Higgs production processes. The impact of pair production is then studied in \autoref{sec:double}. The main results are
summarized and discussed in \autoref{sec:sum} for the most relevant benchmark scenarios considered in the analysis. The appendices collect
some useful formulae and provide additional results for some secondary benchmark scenarios not included in the main text.
Additional numerical results are provided as ancillary files together with the arxiv submission of this paper.

%%%%%%%%%%%%%%%%%%%%%%%%%%%%%%%%%%%%%%%%%%%%%%%%%%%%%%%%%%%%%%%%%%

\section{Low-energy lepton machines}
\label{sec:single}

In this section, we study the precision reach on the trilinear Higgs coupling through the exploitation of single Higgs production measurements. These are the dominant handles available at future circular lepton colliders, like the CEPC and FCC-ee, which cannot easily deliver high luminosities at center-of-mass energies where the Higgs pair production rate becomes sizable. These machines could run above the $\eezhh$ threshold, at a $350\,$GeV center-of-mass energy in particular, but the small cross section (in the attobarn range) and the limited integrated luminosity lead to a negligible sensitivity to this channel. The analysis of single-Higgs production can also be relevant for the ILC. While this machine could eventually reach a center-of-mass energies of $500\,$GeV (or even of $1\,$TeV) in a staged development, its initial low-energy runs can have an impact on the determination of the trilinear Higgs coupling that is worth investigating.

According to recent reports~\cite{CEPCupdate2, FCCupdate}, both CEPC and FCC-ee are planned to collect $5\inab$ of integrated luminosity at $240\,$GeV. FCC-ee is also envisioned to collect $1.5\inab$ at $350\,$GeV.\footnote{The current run plan for FCC-ee anticipates to collect $0.2\inab$ at $350\,$GeV and $1.5\inab$ at $365\,$GeV~\cite{Janot:2017}. Since the vector boson production cross section raises rapidly with the center-of-mass energy, the sensitivity of the FCC-ee will be certainly improved.} Although a run at this center-of-mass energy is not officially forecast for the CEPC, it is nevertheless a viable option given its planned tunnel circumference of $100\,$km. As a general circular collider run scenario, we therefore consider the collection of $5\inab$ of integrated luminosity at $240\,$GeV and several benchmark luminosities at $350\,$GeV, namely $0$, $200\infb$ and $1.5\inab$.

The full ILC run plan comprises $2\inab$ of integrated luminosity at $250\,$GeV, $200\infb$ at $350\,$GeV, and $4\inab$ at $500\,$GeV, with these luminosities equally shared between runs with two $P(e^-, e^+) = (\pm0.8,\mp0.3)$ beam polarization configurations~\cite{Fujii:2015jha, Barklow:2015tja}. Additional results for a $70\%/30\%$ repartition of the luminosity between the $P(e^-, e^+) = (\pm0.8,\mp0.3)$ polarizations will be provided in \autoref{app:results}. 
In this section, we focus only on the runs at $240/250\,$GeV and $350\,$GeV, and consider a few benchmarks for the integrated luminosity collected at $350\,$GeV.

To summarize, we focus on the following benchmark scenarios:
\begin{list}{$\bullet$}{\topsep4pt}

    \item {\bf Circular colliders (CC)} with $5\inab$ at $240\,$GeV, $\{0,~200\infb,~1.5\inab\}$ at $350\,$GeV and unpolarized beams. The scenario with only a $240\,$GeV\,($5\inab$) run corresponds to the {\bf CEPC} Higgs program, while the $240\,$GeV\,($5\inab$)$\,+\,350\,$GeV\,($1.5\inab$) scenario corresponds to the {\bf FCC-ee} Higgs and top-quark programs.
    
    \item {\bf  Low-energy ILC} with $2\inab$ at 250\,GeV, $\{0,~200\infb,~1.5\inab\}$ at 350\,GeV, and integrated luminosities equally shared between $P(e^-, e^+) = (\pm0.8,\mp0.3)$ beam polarizations.\footnote{The current run plan of CLIC anticipates a low-energy operation at $380\,$GeV as a Higgs factory. We did not consider this run alone as the lack of a separate run at a lower energy will constitute an hindrance to the indirect determination of the Higgs cubic self-interaction.}
\end{list}
Later in this section we also extend these scenarios to cover a continuous range of luminosities at $240$\,($250$) and $350$\,GeV.

%%%%%%%%%%%%%%%%%%%%%%%%%%%%%%%%%%%%%%%%%%%%%%%%%%%%%%%%

\subsection{Higher-order corrections to single-Higgs processes}
\label{sec:h3loop}

\begin{figure}[t]
\centering
\includegraphics[width=.6\textwidth]{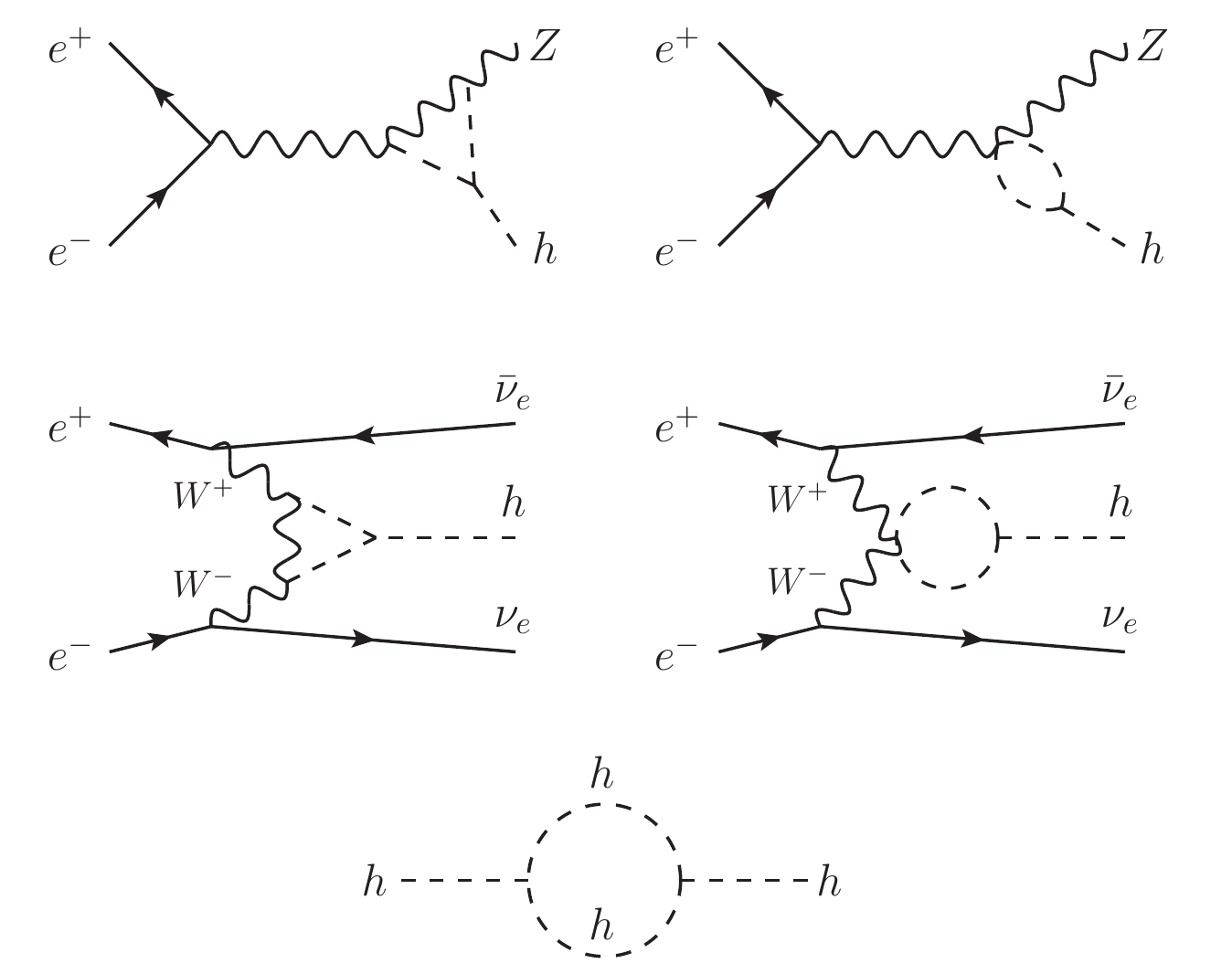}
\caption{One-loop diagrams involving the trilinear Higgs coupling contributing to the main single Higgs production processes: $\eehz$ (top row) and $\eevvh$ (middle row).
The Higgs self-energy diagram (bottom) gives a universal modification to all Higgs production processes via wave function renormalization.} 
\label{fig:feyn1h}
\end{figure}

As a first step, we analyze how a modification of the trilinear Higgs coupling affects single-Higgs processes.
We parametrize possible new physics effects through the quantity $\kappa_\lambda$
defined as the ratio between the actual value of the trilinear Higgs coupling $\lambda_3$ and its SM expression
$\lambda_3^{\rm SM}$ (the Higgs vacuum expectation value is normalized to $v=1/(\sqrt{2}G_{\rm F})^{1/2}\approx 246\,$GeV),\footnote{This parametrization is equivalent to an EFT description in which deviations in the
Higgs trilinear self-coupling arise from a dimension-6 operator $|H^\dagger H|^3$.}
\begin{equation}
\kappa_\lambda \equiv \frac{\lambda_3}{\lambda_3^{\textsc{sm}}}\,, \qquad \lambda_3^{\textsc{sm}} = \frac{m_h^2}{2 v^2}\,.
\end{equation}
While the trilinear coupling does not enter single-Higgs processes at leading order (LO), it affects both Higgs production and decay at next-to-leading order (NLO).
The corresponding diagrams for Higgsstrahlung ($\eehz$) and $WW$-fusion ($\eevvh$) production processes are shown in \autoref{fig:feyn1h}.
In addition to the vertex corrections, which are linear in $\kl$, the trilinear coupling also generates corrections quadratic
in $\kl$ through the wave function renormalization induced by the Higgs self-energy diagram.
Such contributions have been computed
for electroweak~\cite{vanderBij:1985ww, Degrassi:2017ucl, Kribs:2017znd} and single-Higgs observables~\cite{McCullough:2013rea, Shen:2015pha, Gorbahn:2016uoy, Degrassi:2016wml, Bizon:2016wgr, Maltoni:2017ims}.

Following Ref.~\cite{Degrassi:2016wml}, we can parametrize the NLO corrections to an observable $\Sigma$ in a process involving a single external Higgs field as
\begin{equation}
\Sigma_\text{NLO} = Z_H \Sigma_\text{LO}(1+\kl C_1)\,,
\label{eq:definec1}
\end{equation}
where $\Sigma_{\text{LO}}$ denotes the LO value, $C_1$ is a process-dependent coefficient that encodes the interference between the NLO amplitudes involving $\kl$ and the LO ones, while $Z_H$ corresponds to the universal resummed wave-function renormalization and is explicitly given by
\begin{equation}
	Z_H = \frac{1}{1-\kl^2 \dzh}\,,
	\qquad\text{with}\qquad
	\dzh = -\frac{9}{16} \frac{G_\mu m^2_H}{\sqrt{2} \pi^2} \left(\frac{2\pi}{3\sqrt{3}} -1 \right) \simeq -0.00154 \,.
\end{equation}
The impact of a deviation $\dkl\equiv\kl-1$ from the SM value of the trilinear Higgs self-coupling is therefore
\begin{equation}
	\delta \Sigma
%	\;\equiv\; \frac{\Sigma_{\rm NLO}}{\Sigma^{\rm SM}_{\rm NLO}}-1
	\;\equiv\; \frac{\Sigma_{\rm NLO}}{\Sigma_{\rm NLO}(\kappa_\lambda=1)}-1
	\;\simeq\; (C_1+2\dzh)\dkl + \dzh \dkl^2
	\,, \label{eq:kl}
\end{equation}
up to subleading corrections of higher orders in $\dzh$ and $C_1$.\footnote{We checked explicitly that the one-loop squared term of order $\dkl^2$ is subdominant compared to the $\dzh \dkl^2$ one.}
The linear approximation in $\dkl$
is usually accurate enough to describe the deviations in single Higgs processes inside the typical constraint range $|\dkl| \lesssim 5$. We will nevertheless use the unexpanded $\delta\Sigma$ expressions throughout this paper to derive numerical results.

\begin{figure}\centering
\includegraphics[width=.4\textwidth]{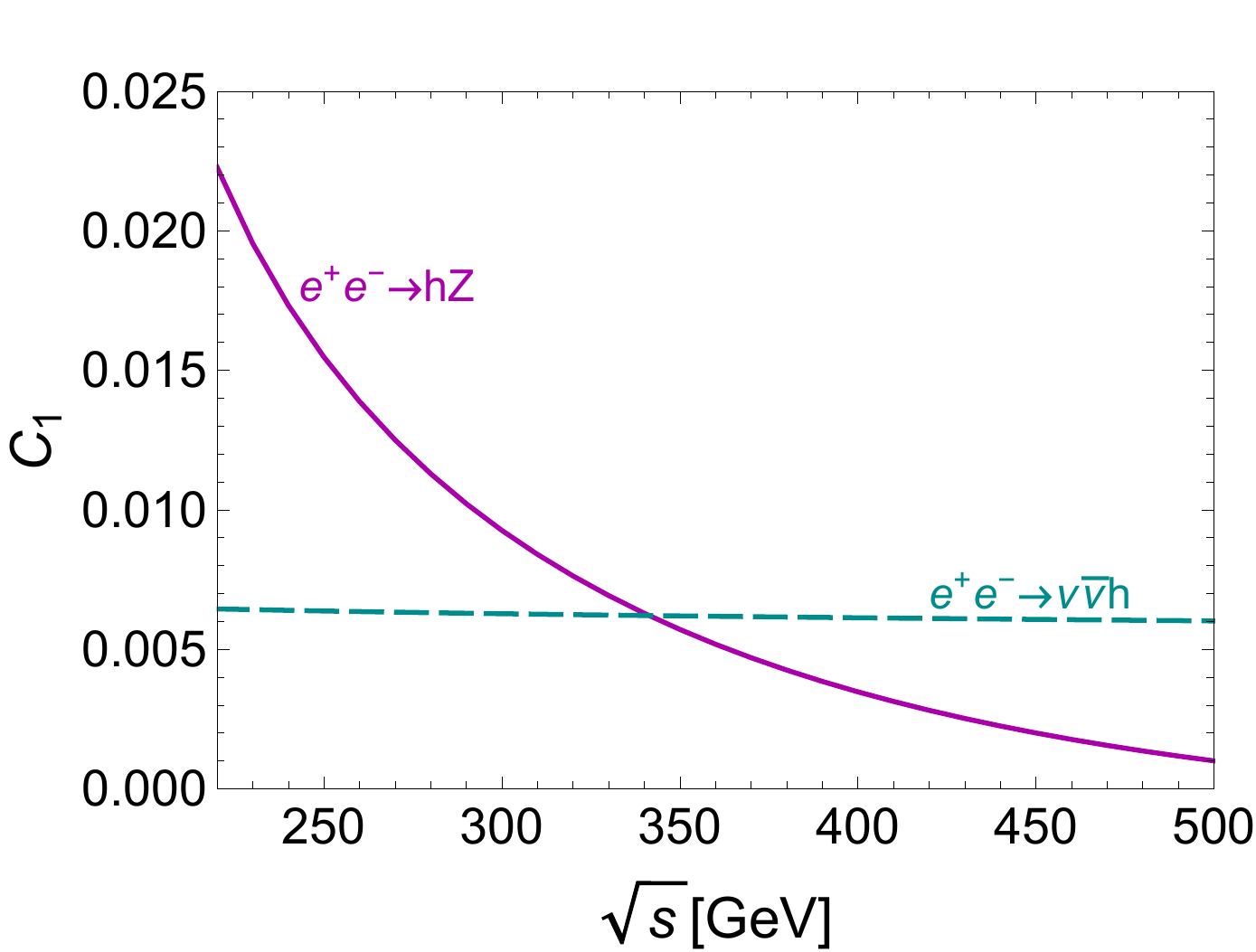} \hfill
\includegraphics[width=.56\textwidth]{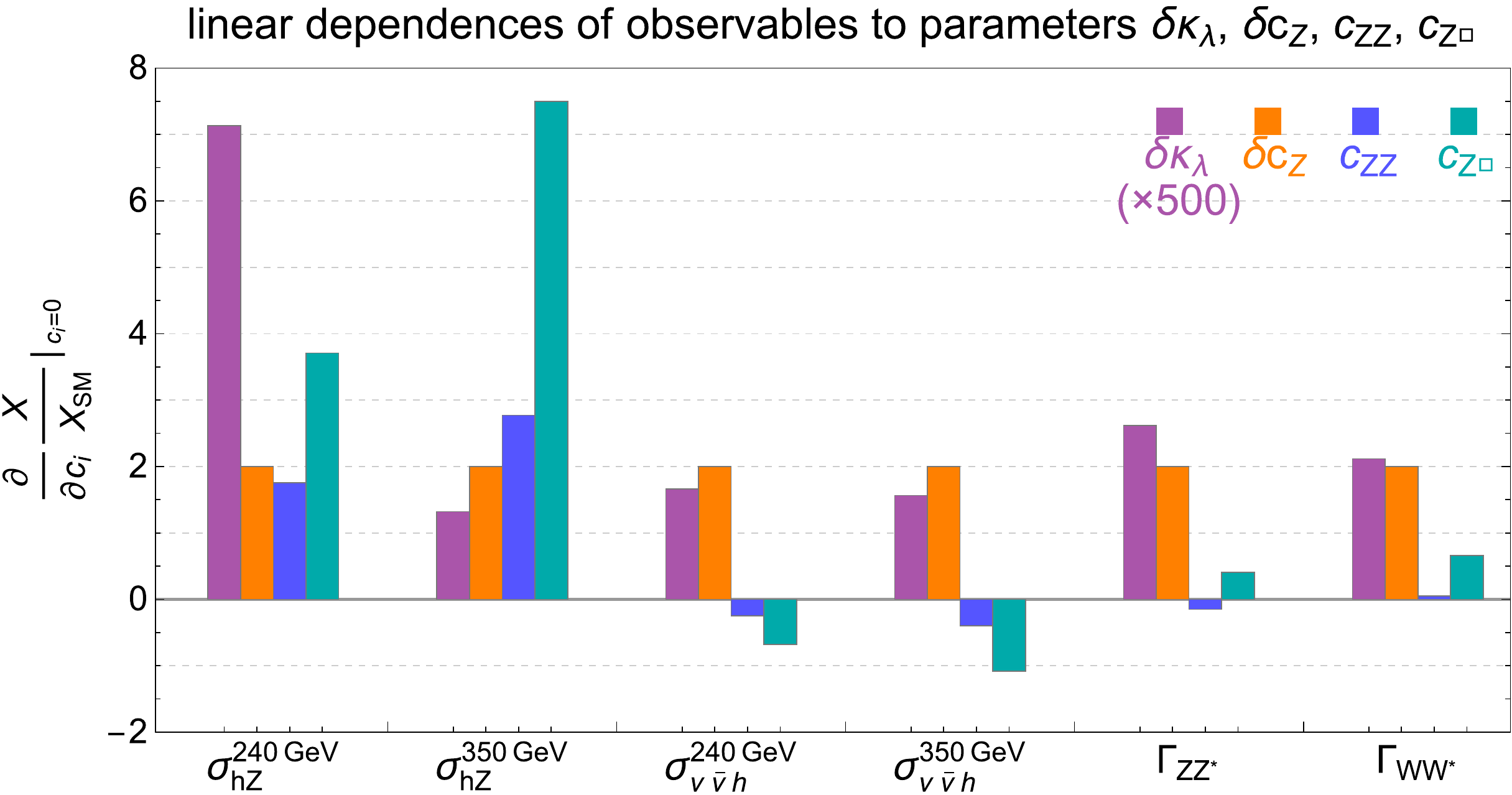}
\caption{{\bf Left:} Value of $C_1$ as a function of the center of mass energy $\sqrt{s}$ for the $\eehz$ and $\eevvh$ single Higgs production processes.
{\bf Right:} The linear dependence of production and decay rates on the $\dkl$, $\delta c_Z$, $c_{ZZ}$ and $c_{Z\square}$ parameters (see \autoref{sec:global1h} for details on the meaning of these parameters). For $\eevvh$, only the $WW$-fusion contribution is included. The dependence on $\dkl$ is amplified by a factor of 500.
}
\label{fig:c1s}
\end{figure}

The value of $C_1$ in Higgsstrahlung ($\eehz$) and $WW$-fusion ($\eevvh$) processes are shown in the left panel of \autoref{fig:c1s} as functions of the center-of-mass energy $\sqrt{s}$. Very different energy dependences are observed for the two processes. A quick decrease is seen in Higgsstrahlung, from $C_1 \simeq 0.022$ at threshold to about $C_1 \simeq 0.001$ at a center-of-mass energy of $500\,$GeV.
On the other hand, a nearly constant value $C_1 \simeq 0.006$ is observed for the $WW$-fusion process over the same range of energy.
Further numerical values are provided in \autoref{app:loop} for both production and decay processes.
Beside the inclusive production and decay rates, we also checked the impact of a correction to $\dkl$
on the angular asymmetries that can be exploited in $\eehz\to h\ell^+\ell^-$ measurements (see Refs.~\cite{Beneke:2014sba, Craig:2015wwr}).
We found that these effects are almost negligible and have no impact on the fits.

To conclude this section, we show in the right panel of \autoref{fig:c1s} the linear dependences of a set of production rates and Higgs partial widths
on $\dkl$ and on three EFT parameters that encode deviations in the $Z$-boson couplings, $\delta c_Z$, $c_{ZZ}$ and $c_{Z\square}$ (see \autoref{sec:global1h} for a detailed discussion of the full set of BSM effects we are considering). Only leading-order dependences are accounted for, at one loop for $\dkl$ and at tree level for the other parameters.
One can see that the various observables have very different dependences on the EFT parameters.
For instance, $\delta c_Z$ affects all the production processes in an energy-independent way.\footnote{In the language of the dimension-six operators, $\delta c_Z$ is generated by the operator $\mathcal{O}_H = \frac{1}{2} (\partial_\mu |H^2| )^2$, which modifies all Higgs couplings universally via the Higgs wave function renormalization.} On the contrary, the effects of $c_{ZZ}$ and $c_{Z\square}$ grow in magnitude for higher center-of-mass energy in both Higgsstrahlung and $WW$-fusion cross sections.
It is apparent that the combination of several measurements can allow us to efficiently disentangle the various BSM effects
and obtain robust constraints on $\dkl$.
From the sensitivities shown in \autoref{fig:c1s}, we can roughly estimate that a set of percent-level measurements in single-Higgs processes has the potential of constraining $\dkl$ with a precision better than ${\cal O}(1)$
and the other Higgs EFT parameters to the percent level.
We will present a detailed quantitative assessment of the achievable precisions in the following.

%%%%%%%%%%%%%%%%%%%%%%%%%%%%%%%%%%%%%%%%%%%%%%%%%%%%%%%%

\subsection{Global analysis}
\label{sec:global1h}

\subsubsection{Analysis of Higgs data at lepton colliders alone} Having obtained the one-loop contributions of $\dkl$ to single Higgs observables, we are now ready to determine the precision reach on the Higgs trilinear self-interaction. In order to obtain a robust estimate, we perform here a global fit, taking into account not only deviations in the Higgs self-coupling, but also corrections to the other SM interactions that can affect single-Higgs production processes.

For our analysis, we follow Ref.~\cite{Durieux:2017rsg}, in which the impact of single-Higgs measurements at lepton colliders on the
determination of Higgs and electroweak parameters was investigated.
We include in the fit the following processes
\begin{list}{$\bullet$}{\itemsep2pt\topsep2pt}
\item Higgsstrahlung production: $e^+ e^- \rightarrow hZ$ (rates and distributions),
\item Higgs production through $WW$-fusion: $e^+ e^- \rightarrow \nu\overline \nu h$,
\item weak boson pair production: $e^+ e^- \rightarrow WW$ (rates and distributions),
\end{list}
with Higgs decaying into a gauge boson pair $ZZ^*$, $WW^*$, $\gamma\gamma$, $Z\gamma$, $gg$ or pairs of fermions $b\overline b$, $c \overline c$, $\tau^+ \tau^-$, $\mu^+ \mu^-$.

New physics effects are parametrized through dimension-six operators within an EFT framework. For definiteness, we express them in the Higgs basis and refer to Ref.~\cite{Falkowski:2001958} for a detailed discussion of the formalism.
Since CP-violating effects are strongly constrained experimentally, we exclusively focus on CP-conserving operators.
We also ignore dipole operators and work under the assumption of flavor universality. We relax this assumption only to consider independent deviations in the of top, bottom, charm, tau, and muon Yukawa couplings.

To estimate the precision in the measurement of the EFT parameters, we assume that the central value of the experimental
results coincides with the SM predictions and we neglect theory uncertainties.
For simplicity we compute the SM cross sections at LO, neglecting NLO effects
coming from SM interactions. These contributions can be important for the experimental analysis, since the modifications they
induce in the SM cross sections can be non negligible compared to the experimental accuracy. For the purpose of estimating
the bounds on BSM effects, however, they play a negligible role.
We adopt a further simplification regarding electroweak precision observables, treating
them as perfectly well measured. Such an assumption can significantly reduce the number of
parameters to consider and is straightforward to implement in the Higgs basis which transparently separates the Higgs and electroweak parameters. The potential impact of this assumption will be discussed at the end of \autoref{sec:sum}.

Under the above assumptions, we are left with twelve independent dimension-six effective operators that can induce leading-order contributions to single-Higgs and diboson processes.  To this set of operators, we add the correction to the Higgs self-coupling parametrized by $\dkl$.\footnote{In the notation of Ref.~\cite{Falkowski:2001958} the $\dkl$ parameter corresponds to $\delta\lambda_3/\lambda$.} The full list of parameters included in our fit contains:
\begin{list}{--}{\itemsep2pt\topsep4pt}
\item corrections to the Higgs couplings to the gauge bosons:
$\delta c_Z$,
$c_{ZZ}$,
$c_{Z\square}$,
$c_{\gamma\gamma}$,
$c_{Z\gamma}$,
$c_{gg}$,
\item corrections to the Yukawa's:
$\delta y_t$,
$\delta y_c$,
$\delta y_b$,
$\delta y_\tau$,
$\delta y_\mu$,
\item corrections to trilinear gauge couplings only:
$\lambda_Z$,
\item correction to the trilinear Higgs self-coupling:
$\dkl$.
\end{list}

Since our focus is on the future sensitivity on the trilinear Higgs self-coupling, we present results in terms of $\dkl$ only, profiling over all other parameters. For a detailed analysis of the sensitivity on the other operators see \autoref{app:results} and Refs.~\cite{Durieux:2017rsg,Barklow:2017awn}.

In our fit, we only include terms linear in the coefficients of the EFT operators, neglecting higher-order corrections.
This approximation can be shown to provide very accurate results for all the parameters entering in our analysis~\cite{Durieux:2017rsg}. The only possible exception is $\dkl$, which can be tested experimentally with much lower precision than the other parameters. Although we checked that a linear approximation is reliable also for $\dkl$,
we keep \autoref{eq:kl} unexpanded in our numerical analyses.
For simplicity, cross terms involving $\dkl$ and other EFT coefficients are however neglected, since the strong constraints on the latter
coefficients and the loop factor make these contributions irrelevant.

In order to estimate the precision on Higgs measurements at different luminosities, we use a naive scaling with
an irreducible $0.1\%$ systematic error. This systematic error has no impact for the benchmark scenarios we consider,
but becomes non-negligible for the large-luminosity projections presented at the end of this section (see \autoref{fig:lulu}).
Another important source of uncertainty in our fit comes from the precision on the determination of trilinear gauge couplings (TGCs). In our analysis, we consider a range of possibilities. In the most conservative case, we assume $1\%$ systematic errors
in each bin of the $e^+ e^- \rightarrow WW$ angular distributions used to constrain anomalous TGCs (aTGCs)~\cite{Durieux:2017rsg}. In the most optimistic
case, we assume that aTGCs are constrained much better than all the other parameters, so that they do not affect our fit. This is equivalent to enforcing the following relations among the EFT parameters:
\begin{align}
\delta g_{1,Z} =&~ \frac{g^2+g'^2}{2(g^2-g'^2)} \left[  -g^2 c_{Z\square} -g'^2c_{ZZ} + e^2 \frac{g'^2}{g^2+g'^2} c_{\gamma\gamma} + g'^2\frac{g^2-g'^2}{g^2+g'^2}c_{Z\gamma}  \right] = 0 \,, \nonumber \\
\delta \kappa_\gamma =&~ -\frac{g^2}{2} \left(  c_{\gamma\gamma} \frac{e^2}{g^2+g'^2} + c_{Z\gamma}\frac{g^2-g'^2}{g^2+g'^2} - c_{ZZ}   \right) = 0\, ,
 \label{eq:tgchb}
\\
\lambda_Z =&~ 0\,.
\nonumber
\end{align}

\begin{figure}[t!]
\centering
\includegraphics[width=.48\textwidth]{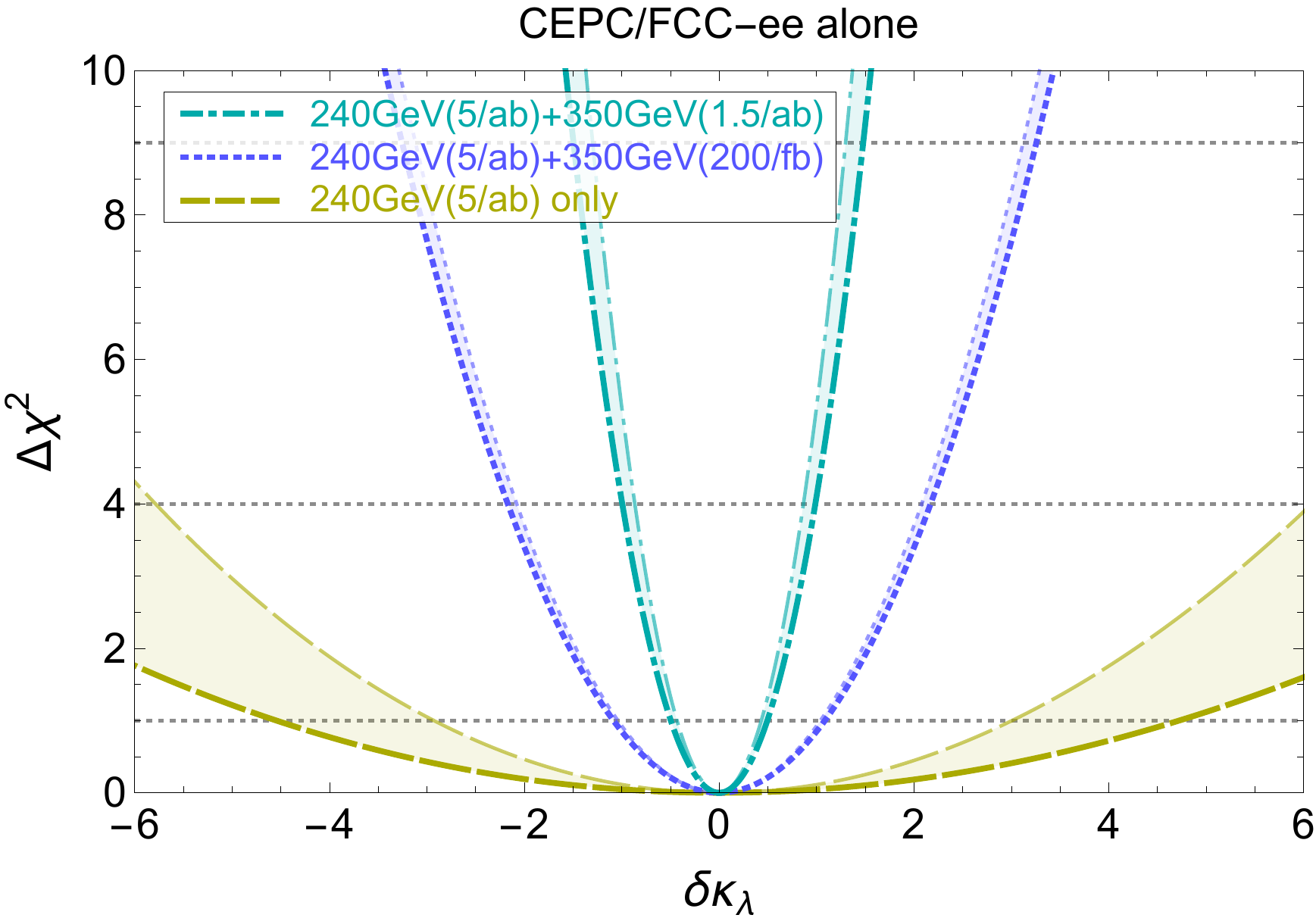}
\includegraphics[width=.48\textwidth]{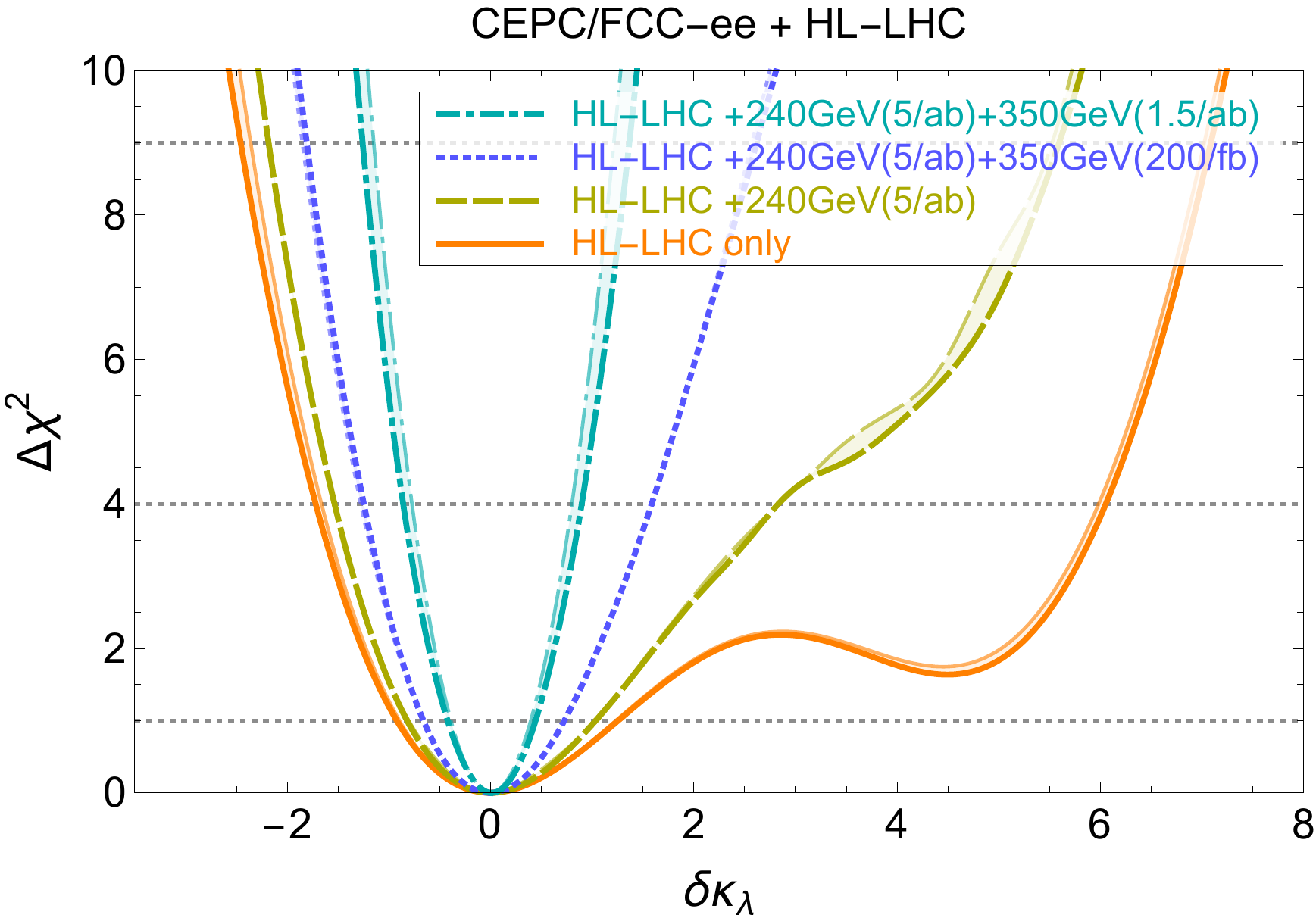}
\caption{Chi-square as a function of $\dkl$ after profiling over all other EFT parameters. Three run scenario are considered for circular colliders, with $5\inab$ at $240\,$GeV and $\{0,~200\infb,~1.5\inab\}$ at $350\,$GeV, without beam polarization. The shaded areas cover different assumptions about the precision of TGC measurements.
{\bf Left:} circular lepton collider measurements only. {\bf Right:} combination with differential single and double Higgs measurements at the HL-LHC.
}
\label{fig:npcepc1}
\end{figure}

\medskip

We start our discussion of the fit results by considering the benchmark scenarios for circular colliders.
The profiled $\Delta \chi^2$ fit as a function of $\dkl$ is shown in the left panel of \autoref{fig:npcepc1}.
The $68\%$\,CL intervals are also reported in \autoref{tab:resultscepc}.

\begin{table}[t]
\centering
\begin{tabular}{@{\hspace{4pt}}l@{\hspace{-2cm}}l||c|c||c|c@{\hspace{4pt}}}
&& \multicolumn{2}{c||}{\small lepton collider alone} & \multicolumn{2}{|c}{\small lepton collider + HL-LHC}\\
\cline{3-6}
\rule[-.4em]{0pt}{1.3em} &&    {\scriptsize non-zero aTGCs} &   {\scriptsize  zero aTGCs} 
  &    {\scriptsize non-zero aTGCs} &  {\scriptsize  zero aTGCs}  \\
\hline \hline
\rule[-.6em]{0pt}{1.7em}{\scriptsize HL-LHC alone}&       &    &     & {\small $[{-0.92},{+1.26}]$} &   {\small $[{-0.90},{+1.24}]$} \\
\hline
\rule[-.6em]{0pt}{1.7em}{\scriptsize CC 240\,GeV ($5\inab$)}&               & {\small $[{-4.55},{+4.72}]$}  & {\small $[{-2.93},{+3.01}]$}  &  {\small $[{-0.81},{+1.04}]$}  &  {\small $[{-0.82},{+1.03}]$}\\
\hline
&\rule[-.6em]{0pt}{1.7em}{\scriptsize +350\,GeV ($200\infb$)}  & {\small $[{-1.08},{+1.09}]$} & {\small $[{-1.04},{+1.04}]$}  &  {\small $[{-0.66},{+0.76}]$}   &  {\small $[{-0.66},{+0.74}]$} \\
\hline
&\rule[-.6em]{0pt}{1.7em}{\scriptsize +350\,GeV ($1.5\inab$)}     & {\small $[{-0.50},{+0.49}]$}   & {\small $[{-0.43},{+0.43}]$} &  {\small $[{-0.43},{+0.44}]$}  &  {\small $[{-0.39},{+0.40}]$} \\ \hline  
\rule[-.6em]{0pt}{1.7em}{\scriptsize ILC 250\,GeV ($2\inab$)}&               & {\small $[{-5.72},{+5.87}]$}  & {\small $[{-5.39},{+5.62}]$}  &  {\small $[{-0.85},{+1.13}]$}  &  {\small $[{-0.85},{+1.12}]$}\\
\hline
&\rule[-.6em]{0pt}{1.7em}{\scriptsize +350\,GeV ($200\infb$) }               & {\small $[{-1.26},{+1.26}]$} & {\small $[{-1.18},{+1.18}]$}  &  {\small $[{-0.72},{+0.83}]$}   &  {\small $[{-0.71},{+0.80}]$} \\
\hline
&\rule[-.6em]{0pt}{1.7em}{\scriptsize +350\,GeV ($1.5\inab$)}               & {\small $[{-0.64},{+0.64}]$}   & {\small $[{-0.56},{+0.56}]$} &  {\small $[{-0.52},{+0.54}]$}  &  {\small $[{-0.48},{+0.50}]$}
\end{tabular}
\caption{One-sigma bounds on $\dkl$ from single-Higgs measurements at circular lepton colliders (denoted as CC) and the ILC.
The first column shows the results for lepton colliders alone, while the second shows the combination with differential measurements of both single and double Higgs processes at the HL-LHC. For each scenario two benchmarks with
conservative and optimistic assumptions on the precision on trilinear gauge couplings are listed. The integrated luminosity is assumed equally shared between $P(e^-,e^+)=(\pm0.8,\mp0.3)$ for the ILC.
}
\label{tab:resultscepc}
\end{table}

The numerical results show that a $240\,$GeV run alone has a very poor discriminating power on the Higgs trilinear coupling, so that
only an ${\cal O}(\textit{few})$ determination is possible (brown dashed lines in the plot). The constraint is also highly sensitive to the precision in the determination of TGCs, as can be inferred from the significantly different bounds in the conservative and optimistic aTGCs scenarios. The inclusion of measurements at $350\,$GeV drastically improve the results. An integrated luminosity
of $200\infb$ at $350\,$GeV, is already sufficient to reduce the uncertainty to the level $|\dkl| \lesssim 1$, whereas $1.5\inab$
leads to a precision $|\dkl| \lesssim 0.5$.

It is interesting to compare the above results with the constraints coming from an exclusive fit in which only corrections
to the trilinear Higgs coupling are considered and all the other parameters are set to zero. With $5\inab$ collected at $240/250\,$GeV, and irrespectively of the presence of a run at $350\,$GeV, we find that such a fit
gives a precision of approximately $14\%$ in the determination of $\dkl$. The strongest constraints come from the measurement
of the $e^+ e^- \rightarrow Zh$ cross section at the $240\,$GeV run, which is the observable with the largest sensitivity to
$\dkl$ (see discussion in \autoref{sec:global1h} and left panel of \autoref{fig:c1s}). Other processes at the $240\,$GeV run and
the higher-energy runs have only a marginal impact on the exclusive fit.

\begin{figure}[t]
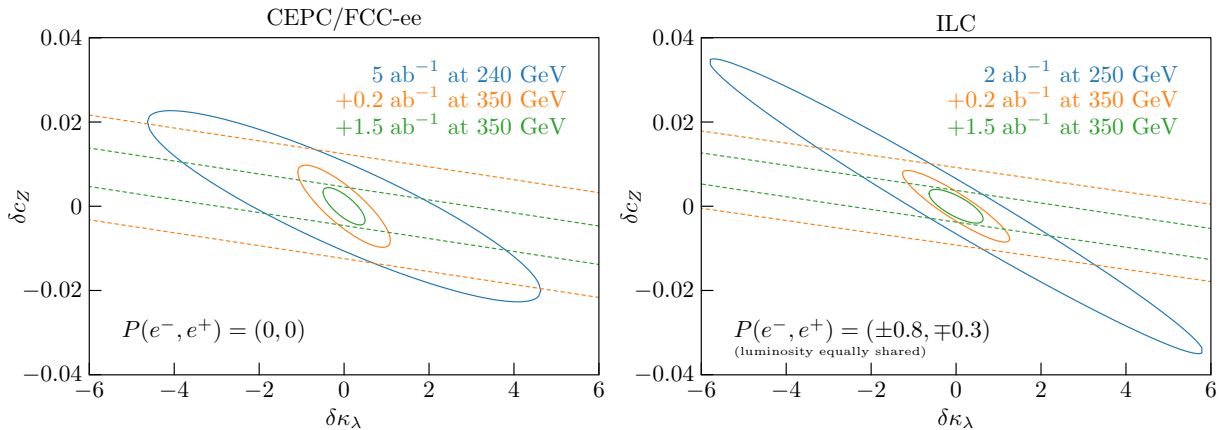

\centering
\includegraphics[scale=.88]{dcz-dkl-cc-2.mps}\hfill
\includegraphics[scale=.88]{dcz-dkl-lc-2.mps}
\caption{Global constraints on $\delta c_Z$ and $\dkl$, obtained from single Higgs measurements at circular colliders (left panel) and ILC (right panel), illustrating the improvement brought by $350\,$GeV runs. Dashed lines are for the latter only, while solid lines combined them with the $240/250\,$GeV one.
}
\label{fig:2d-fit}
\end{figure}

The exclusive fit provides a bound much stronger than the
global analyses, signaling the presence of a nearly flat direction in the global fits.
We found that $\dkl$ has a strong correlation
with $\delta c_Z$ and $c_{gg}$, while milder correlations are present with $c_{Z\square}$ and $\lambda_Z$.\footnote{Notice that a loosely constrained direction
involving $\delta c_Z$ is already present in the global fit not including $\dkl$~\cite{Durieux:2017rsg}. The addition of the trilinear Higgs coupling
makes this feature even more prominent.}
This result sheds some light on the origin of the improvement in the global fit coming from the combination of the $240\,$GeV
and $350\,$GeV runs. The latter runs, although probing processes with a smaller direct sensitivity to $\dkl$, are
useful to reduce the uncertainty on the other EFT parameters. In particular, the $350\,$GeV run with $1.5\inab$ of integrated luminosity
allows for a reduction of the uncertainty on $\delta c_Z$, $c_{gg}$, $c_{Z\square}$ and $\lambda_Z$ by a factor of about $4$.
This in turn helps in lifting the flat direction in the global fit. This effect is clearly visible from the left panel of \autoref{fig:2d-fit},
which shows the fit on the $\dkl$ and $\delta c_Z$ parameters obtained with a $240\,$GeV run only and with the inclusion of
a $350\,$GeV run.

\subsubsection{Synergy between measurements at the HL-LHC and lepton-colliders}
So far, we only considered the precision reach of lepton colliders on the extraction of the trilinear Higgs self-coupling.
Significant information on $\dkl$ can however also be obtained at the high-luminosity LHC. It is thus interesting
to estimate the impact of combining the different sets of measurements.

The Higgs trilinear self-coupling can be accessed at the HL-LHC mainly through the exploitation of the Higgs pair
production channel $p p \rightarrow h h$. An analysis of this channel within the EFT framework has been presented in
Ref.~\cite{Azatov:2015oxa}, in which the most promising channel, namely $p p \rightarrow hh \rightarrow b \overline b \gamma\gamma$, has been investigated.
A fully differential analysis (taking into account the Higgs pair invariant mass distribution) allows to constrain $\dkl$
to the interval $[-1.0, 1.8]$ at $68\%$ CL. A second minimum is however present in the fit, which allows for sizable
positive deviations in $\dkl$, namely an additional interval $\dkl \in [3.5, 5.1]$ can not be excluded at $68\%$ CL.
Some improvement can be obtained complementing the pair-production channel
with information from single Higgs channels, which are affected at NLO by the Higgs self-coupling. In this way,
the overall precision becomes $\dkl \in [-0.9, 1.2]$ at $68\%$ CL (with the additional minimum at $\dkl \sim 5$ excluded) and $\dkl \in [-1.7, 6.1]$ at $95\%$ CL~\cite{DiVita:2017eyz}.
To estimate the impact of HL-LHC, we will use here the results of the combined fit with differential single and pair production
(corresponding to the orange solid curve in the right panel of \autoref{fig:npcepc1}).

The combinations of the HL-LHC fit with our benchmarks for circular lepton colliders are shown in the right panel of
\autoref{fig:npcepc1}. One can see that a $240\,$GeV run is already sufficient to completely lift the second minimum at
$\dkl \sim 5$, thus significantly reducing the $2\sigma$ bounds. The precision near the SM point ($\dkl = 0$) is however
dominated by the HL-LHC measurements, so that the lepton collider data can only marginally improve the $1\sigma$ bounds.
The situation is reversed for the benchmarks including a $350\,$GeV run. In this case, the precision achievable at lepton
colliders is significantly better than the HL-LHC one.
The combination of the LHC and lepton collider data can still allow for a significant improvement in the constraints if limited integrated luminosity can be accumulated in the $350\,$GeV runs (see \autoref{tab:resultscepc}). With
$1.5\inab$ collected at $350\,$GeV, on the other hand, the lepton collider measurements completely dominate the bounds.

\medskip

Similar results are obtained for the low-energy ILC benchmarks. In this case, the lower integrated luminosity forecast at
$250\,$GeV ($2\inab$) can be compensated through the exploitation of the two different beam polarizations
$P(e^-, e^+) = (\pm 0.8, \mp 0.3)$. The only difference with respect to the circular collider case is the fact that
the $250\,$GeV run fit is more stable under changes in the trilinear gauge couplings precision. This is due to the
availability of runs with different polarizations, which provide better constraints on the EFT parameters.
Analogously to the circular collider scenarios, the combination of the $250\,$GeV
measurements with the HL-LHC data allows to completely lift the minimum at $\dkl \sim 5$, while a $350\,$GeV run
would easily surpass the LHC precision. We report the results for the ILC benchmarks in \autoref{app:results} (see \autoref{fig:npilc1}). For completeness, we mention that an exclusive fit on $\dkl$ at the ILC allows for a precision
of approximately $32\%$, significantly better than the one expected through a global fit. Also in this case a nearly flat direction
is present when deviations in all the EFT parameters are simultaneously allowed (see right panel of \autoref{fig:2d-fit}).

\medskip

\begin{figure}[t]
\centering
\includegraphics[width=.48\textwidth]{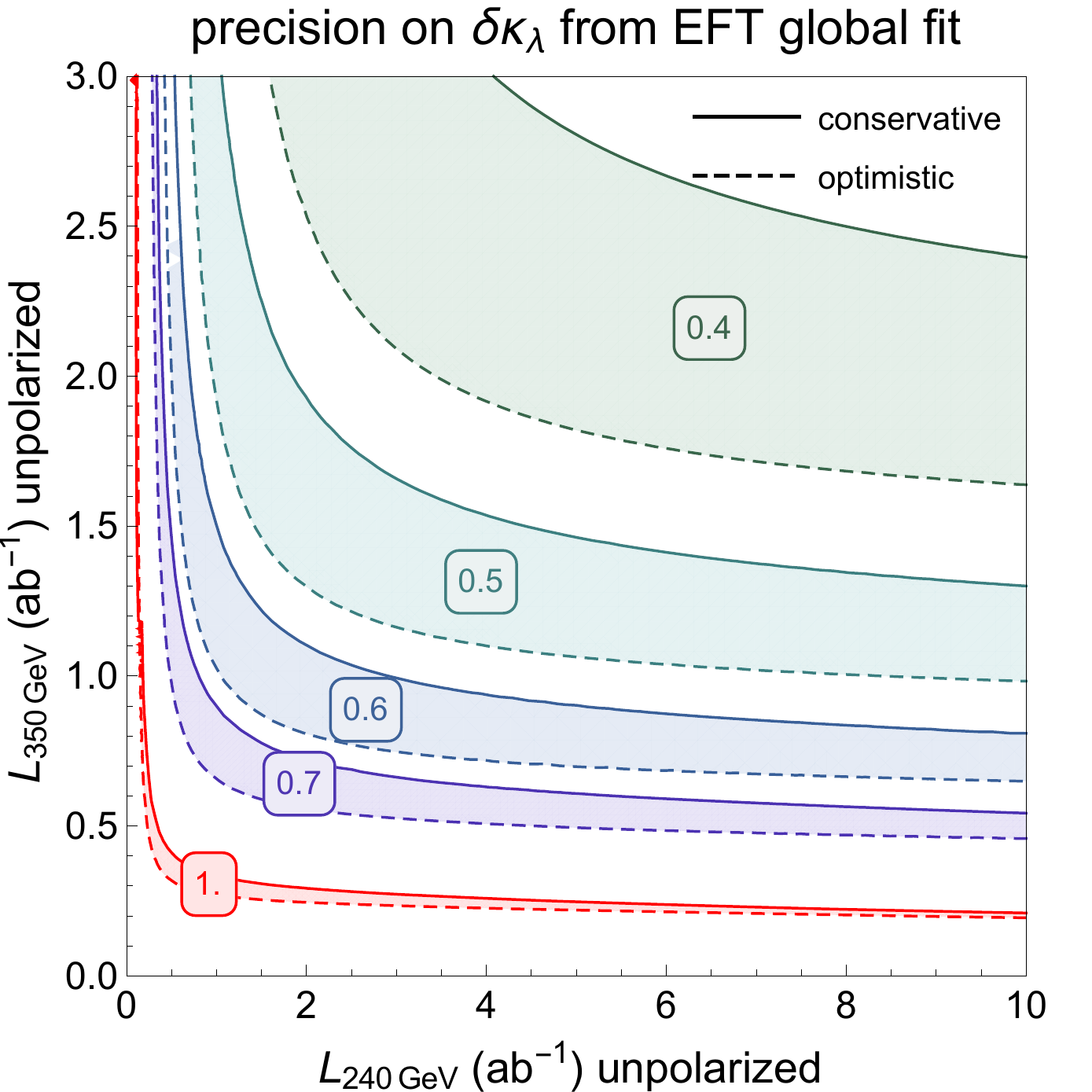}\hfill
\includegraphics[width=.48\textwidth]{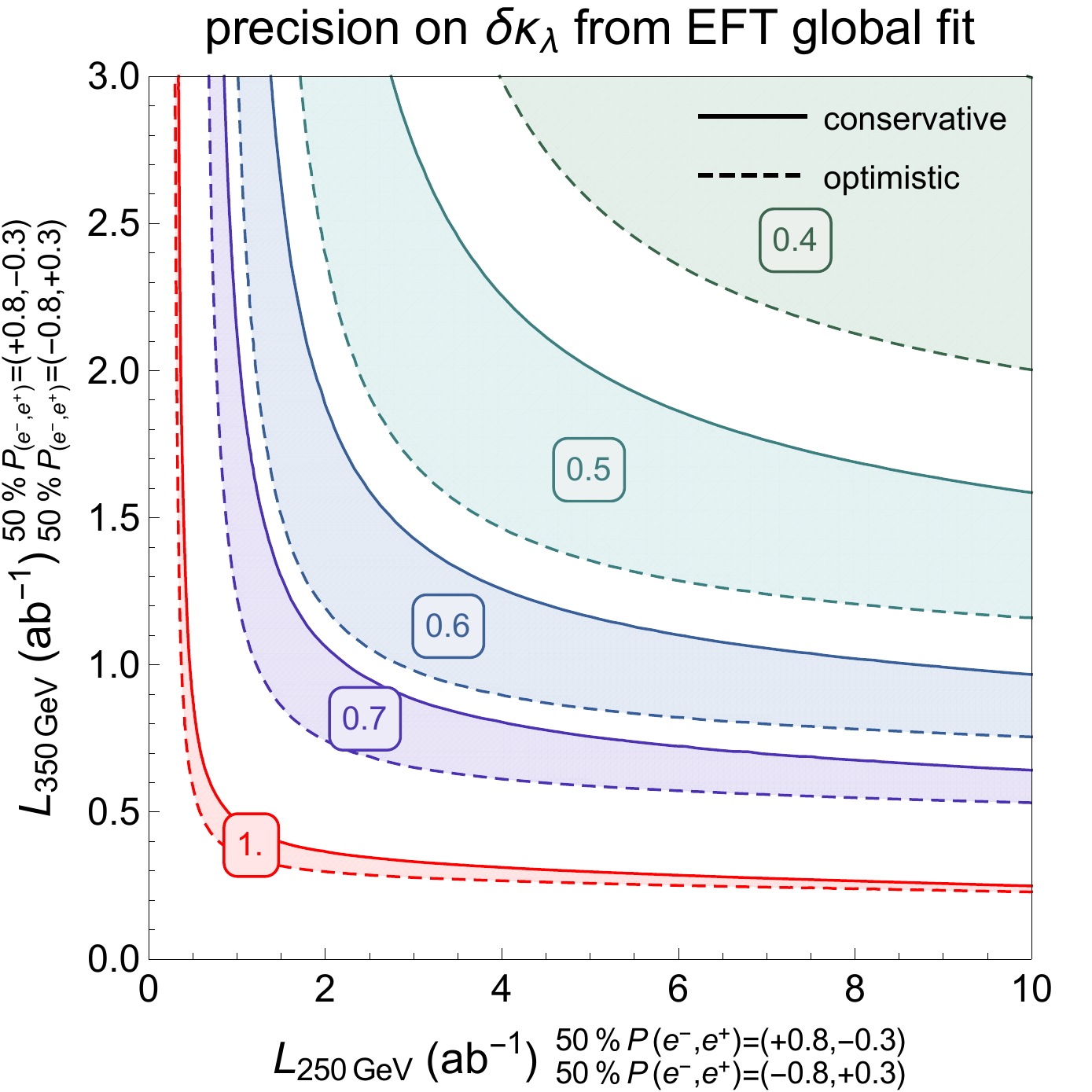}
\caption{One-sigma bound on $\dkl$ deriving from single Higgs and diboson production measurements at lepton colliders as a function of the integrated luminosity collected at both $240/250$ and $350\,$GeV. Conservative (solid) and optimistic (dashed) assumptions are used for the precision of diboson measurements.}
\label{fig:lulu}
\end{figure}

Having observed the significant impact of the combination of measurements at  $240/250$ GeV and $350\,$GeV center-of-mass energies, to conclude the discussion, we now explore a continuous range of integrated luminosities accumulated at the various colliders.
The one-sigma limits as functions of the integrated luminosity are displayed in \autoref{fig:lulu} for the circular colliders and the ILC.
Conservative and optimistic precisions for TGC measurements are respectively assumed to obtain the solid and dashed curves.
The combination of runs at these two different energies always brings drastic improvements. The fastest improvements in precision on the $\dkl$ determination is obtained along the $L_\text{350\,{\rm GeV}}/L_\text{240\,{\rm GeV}}\simeq 0.7$ and $L_\text{350\,{\rm GeV}}/L_\text{250\,{\rm GeV}} \simeq 0.5$ lines for circular colliders and the ILC, respectively.

\section{High-energy lepton machines}
\label{sec:double}

Having explored the reach of low-energy lepton colliders in the previous section, we now enlarge our scope
to include machines with center-of-mass energies above $350\,$GeV. They offer the opportunity of probing directly the trilinear Higgs self-coupling through
Higgs pair production processes, double Higgsstrahlung $\eezhh$ and $WW$-fusion $\eevvhh$ in particular.
The precision reach in the determination of $\dkl$ at ILC and CLIC has already been studied by the
experimental collaborations~\cite{Duerig:2016dvi,Abramowicz:2016zbo}. These studies performed an exclusive fit,
allowing for new-physics effects only in the trilinear Higgs self-coupling.

In this section, we first review the experimental projections on the extraction of the Higgs self-coupling
from double Higgs channels. In this context, we also point out how differential distributions, in particular
in the $WW$-fusion channel, can allow for an enhanced sensitivity to $\dkl$.
Afterwards, we reconsider Higgs pair production measurements from a global EFT perspective, showing how the
determination of $\dkl$ is modified by performing a simultaneous fit for all EFT parameters. We also evaluate
how these results are modified by combining double-Higgs data with single-Higgs measurements from low-energy runs.

%%%%%%%%%%%%%%%%%%%%%%%%%%%%%%%%%%%%%%%%%%%%%%%%%%%%%%%%%%%%%%%%%%

\subsection{Higgs pair production}
\label{sec:exp2h}

\begin{figure}[tb]\centering
\includegraphics[width=.48\textwidth]{hh_xsec_old.mps}
\hfill
\raisebox{1em}{\includegraphics[width=.50\textwidth]{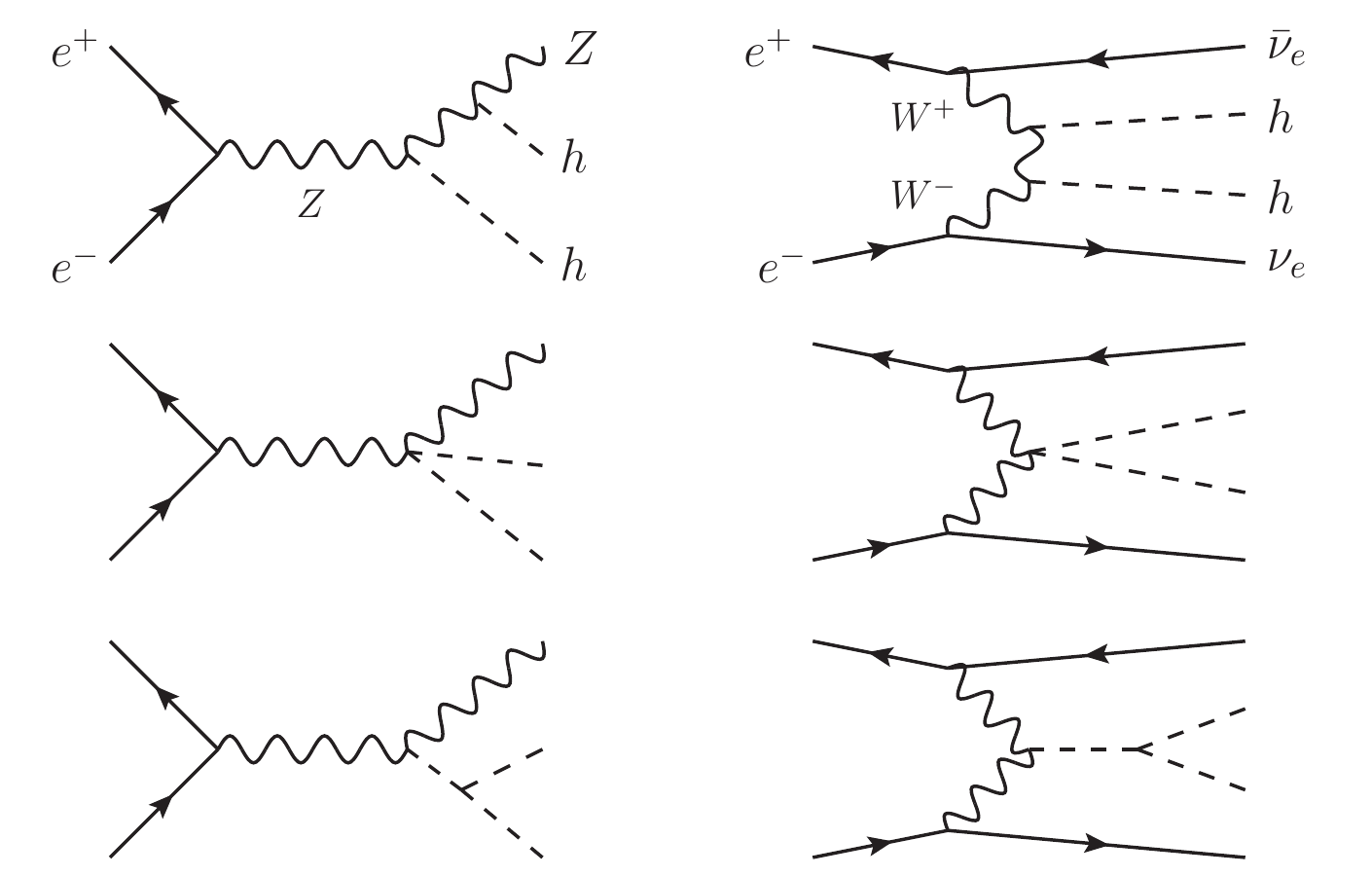}}
\caption{Higgs pair production cross sections at lepton colliders as functions of the center-of-mass energy (based on Fig.\,7 of Ref.\,\cite{Tian:2013yda}) and illustrative diagrams. The difference between the two $\vvhh$ curves is entirely due to double Higgsstrahlung followed by invisible $Z$ decay.
}
\label{fig:hh_xsec}
\end{figure}

As already mentioned, Higgs pair production at high-energy lepton machines is accessible mainly through the double Higgsstrahlung $\eezhh$ and $WW$-fusion $\eevvhh$ channels. The cross sections for these two production modes
as functions of the center-of-mass energy of the collider are shown in \autoref{fig:hh_xsec}. It is interesting to notice their completely different behavior,
so that the relevance of the two channels drastically changes at different machines. At energies below approximately $1\,$TeV,
double Higgsstrahlung is dominant whereas, at higher energy, the channel with the larger cross section is $WW$-fusion.
To be more specific, the cross section of double Higgsstrahlung reaches a maximum at $\sqrt{s} \simeq 600\,$GeV
before starting to slowly decrease as the $s$-channel $Z$ boson gets more and more offshell.
On the contrary, the $e^+e^-\to\vvhh$ cross section initially grows steadily with the center-of-mass energy of the collider and
adopts a logarithmic behavior above $10\,$TeV.
Notice that the $e^+e^-\to\vvhh$ channel receives non-negligible contributions
that are not of $WW$-fusion type. The largest of them arises from double Higgsstrahlung followed by a $Z\to \nu\bar\nu$ decay. These contributions can however be efficiently identified at sufficiently
high center-of-mass energies since the kinematic of the process is significantly different from that of $WW$-fusion.
Notice, moreover, that both double-Higgs production cross sections are significantly affected by the beam polarization (see \autoref{app:results} and \autoref{fig:hh_xsec2}).

The $\eezhh$ process at the ILC with $500\,$GeV center-of-mass energy has been thoroughly studied in Ref.~\cite{Duerig:2016dvi}. A total luminosity of $4\inab$, equally split into two beam polarization runs  $P(e^-,e^+)=(\pm 0.8, \mp 0.3)$, allows for a precision of $21.1\%$ on the cross section determination through the exploitation of the $hh \to b\bar{b}b\bar{b}$ final state. A further improvement can be obtained by also including the $hh \to b\bar{b}WW^*$ channel, in which case the precision reaches $16.8\%$.

The $\eevvhh$ process has also been studied at a $1\,$TeV center-of-mass energy. A significance of $2.7\sigma$ (corresponding to a precision of $37\%$) could be achieved in the $hh \to b\bar{b}b\bar{b}$ channel, assuming and integrated luminosity $\mathcal{L}=2\inab$ and $P(e^-,e^+)=(- 0.8, + 0.2)$ beam polarization~\cite{ILC1TeV}.

Studies of the $\eevvhh$ process at CLIC (both at $1.4\,$TeV and $3\,$TeV center-of-mass energy) are
available in Ref.~\cite{Abramowicz:2016zbo}.
Assuming unpolarized beams and $1.5\inab$, the precision on the $1.4\,$TeV cross section could reach $44\%$.
With $1.5\inab$, the $3\,$TeV cross section could be measured with a $20\%$ precision. Both $b\bar{b}b\bar{b}$ and $b\bar{b}WW^*$ channels are included in these analyses, though the sensitivity is mainly driven by the former,
as shown in Table~28 in Ref.~\cite{Abramowicz:2016zbo}.

\begin{figure}[tb]
\centering
\includegraphics[width=.467\textwidth]{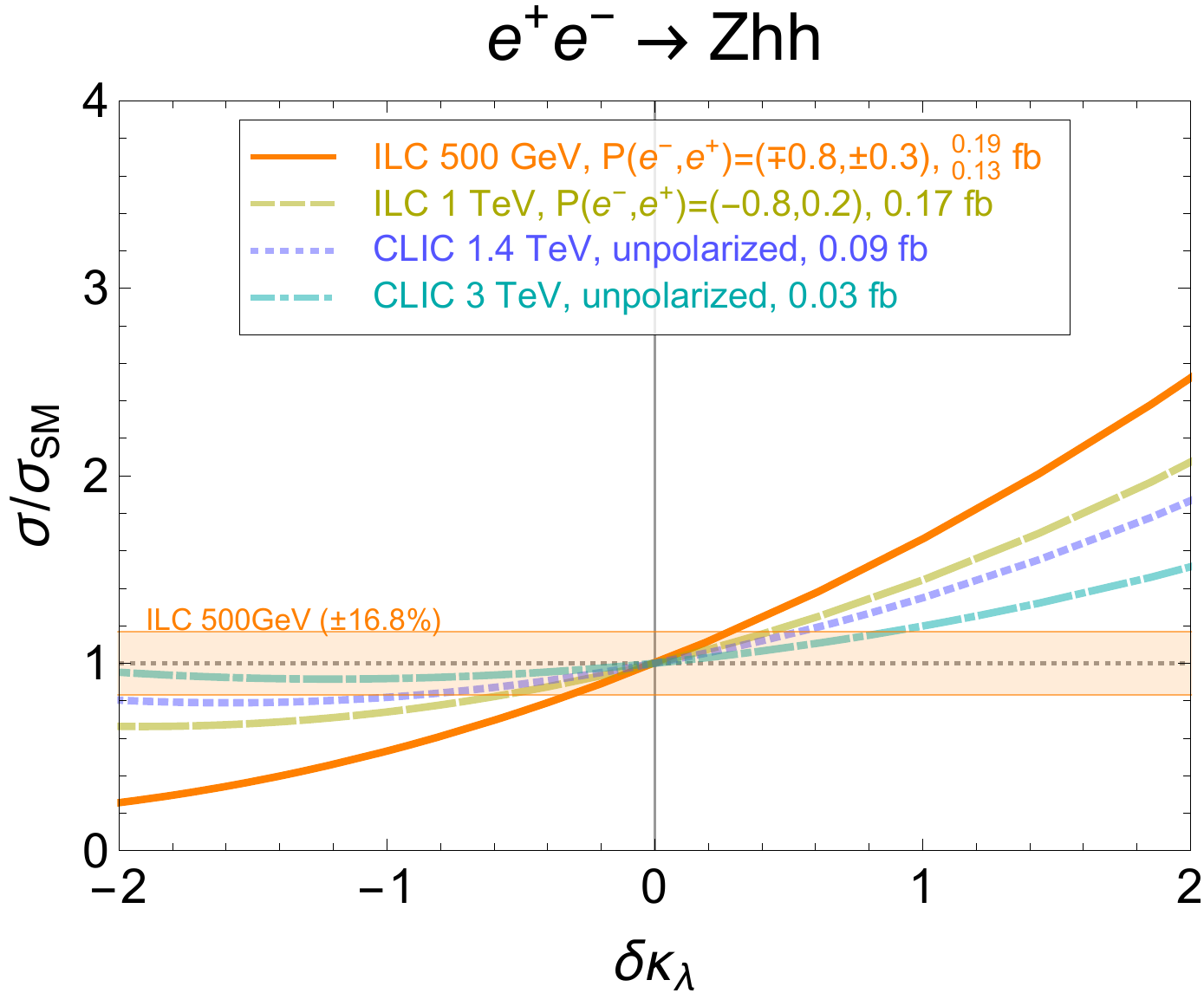} \hspace{0.1cm}
\includegraphics[width=.475\textwidth]{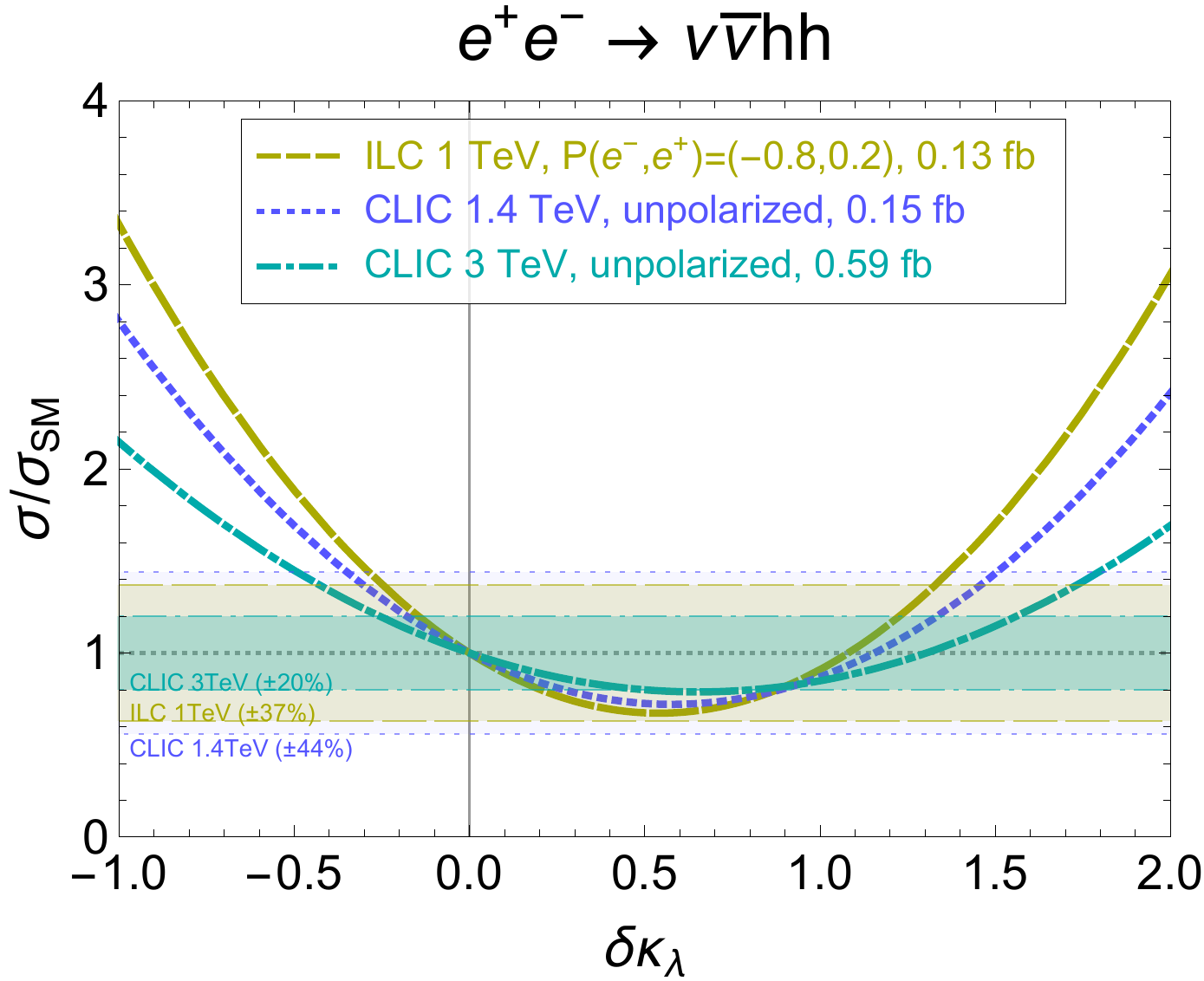}
\caption{Dependence of the Higgs pair production rates on $\dkl$ at various center-of-mass energies. Shaded bands display the precision claimed by dedicated experimental studies on the standard-model cross sections. Absolute cross sections are provided in the legend.
}
\label{fig:hhsen1}
\end{figure}

\begin{table}[h!]
\centering
\begin{tabular}{c|c|c}
& 68 \%CL & 95\%CL\\
\hline
\hline
\rule[-.5em]{0pt}{1.6em}ILC 500\,GeV   &  $[-0.31,~0.28]$    &  $[-0.67,~0.54]$    \\ \hline
\rule[-.5em]{0pt}{1.6em}ILC 1\,TeV   &  $[-0.25,~1.33]$    &  $[-0.44,~1.52]$    \\ \hline
\rule[-.5em]{0pt}{1.6em}ILC combined   &  $[-0.20,~0.23]$    &  $[-0.37,~0.49]$    \\ \hline\hline
\rule[-.5em]{0pt}{1.6em}CLIC 1.4\,TeV   &  $[-0.35,~1.51]$    &  $[-0.60,~1.76]$    \\ \hline
\rule[-.5em]{0pt}{1.6em}CLIC 3\,TeV   & $[-0.26,~0.50] \cup [0.81,~1.56]$     &  $[-0.46,~1.76]$   \\ \hline
\rule[-.5em]{0pt}{1.6em}CLIC combined   &  $[-0.22,~0.36] \cup [0.90,~1.46]$     &  $[-0.39,~1.63]$    \\ \hline
\rule[-.5em]{0pt}{1.6em} $+Zhh$   &  $[-0.22,~0.34] \cup [1.07,~1.28]$     &  $[-0.39,~1.56]$    \\ \hline
\rule[-.5em]{0pt}{1.6em} 2 bins in $\vvhh$   &  $[-0.19,~0.31]$   &  $[-0.33,~1.23]$    \\ \hline
\rule[-.5em]{0pt}{1.6em} 4 bins in $\vvhh$   &  $[-0.18,~0.30]$   &  $[-0.33,~1.11]$
\end{tabular}
\caption{
Constraints from an exclusive fit on $\dkl$ derived from the measurements of $\vvhh$ and $\eevvhh$ cross sections
at ILC and CLIC, with all other parameters fixed to their standard-model values.
}
\label{tab:clicvvhh01}
\end{table}

\medskip

The dependence of the Higgs pair production cross sections on $\dkl$ is shown in \autoref{fig:hhsen1} for a set of benchmark scenarios.
The SM cross section for each benchmark is provided in the legend.\footnote{The ILC 1\,TeV SM cross section is obtained from Fig.\,7 of Ref.\,\cite{Tian:2013yda} and scaled from $P(e^-,e^+)=(-0.8,+0.3)$ to $P(e^-,e^+)=(-0.8,+0.2)$. The unpolarized CLIC SM cross sections are taken from Ref.~\cite{Abramowicz:2016zbo}.}
Shaded bands show the precisions on the determination of the SM rates discussed above.
Note the experimental collaborations made no forecast for the precision on double Higgsstrahlung at $1\,$TeV and above.

It is interesting to notice that, around the SM point, the sensitivity of both Higgs pair production channels to $\dkl$ gets milder at higher center-of-mass energy. On the contrary, the sensitivity to the other EFT parameters tends to increase with energy.
Another important feature is the significant impact of terms quadratic in $\dkl$ on the behavior of the cross section around the SM point, especially for the $WW$-fusion channel shown in the right panel of \autoref{fig:hhsen1}. For this reason, a linear
approximation is in many cases not sufficient to extract reliable bounds. In \autoref{tab:clicvvhh01}, we list the $68\%$ and $95\%$ CL bounds obtained from the benchmarks ILC and CLIC runs retaining the
full dependence of the cross section on $\dkl$.

\begin{figure}[t]
\centering
\includegraphics[width=.48\textwidth]{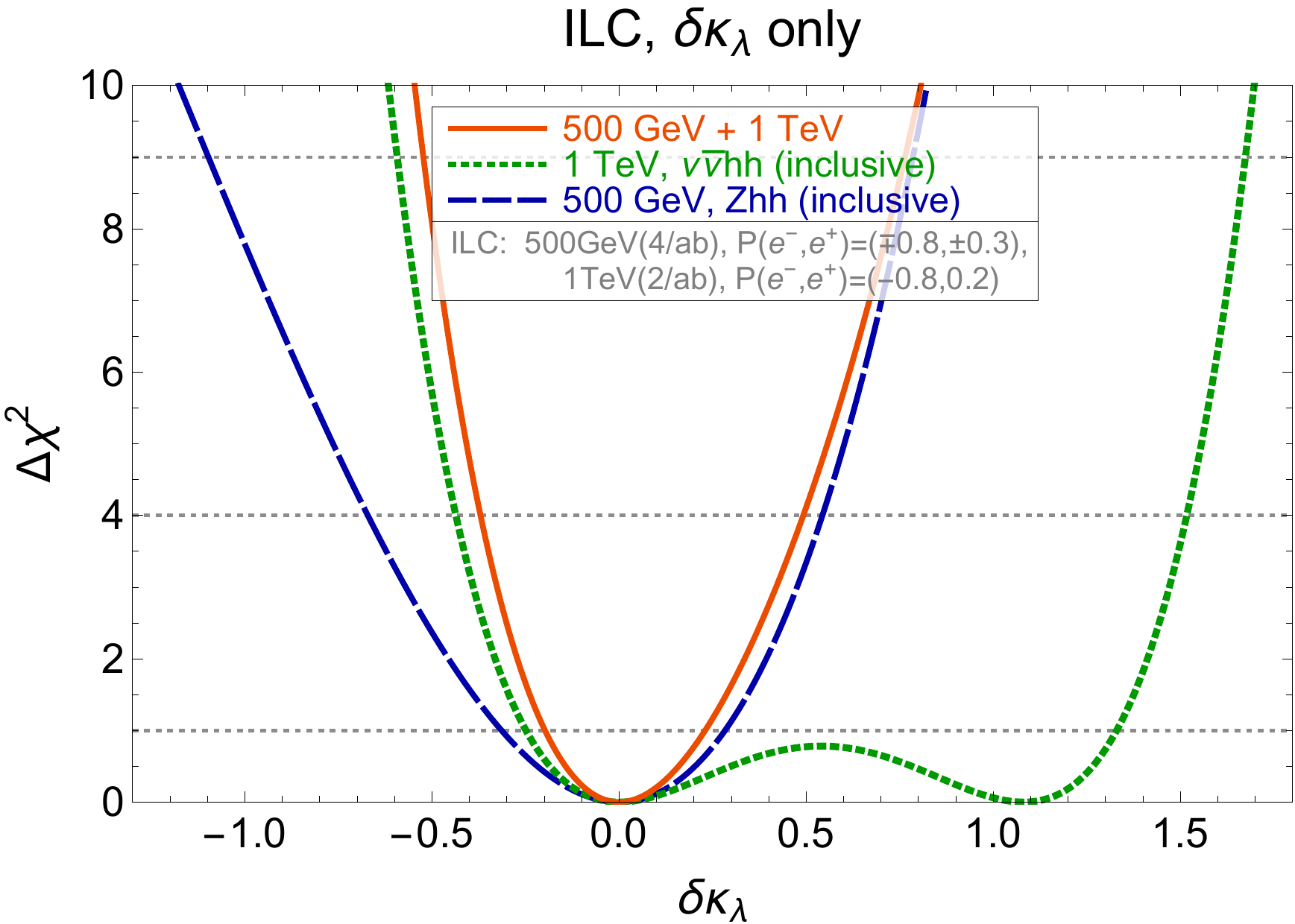}
\hfill
\includegraphics[width=.48\textwidth]{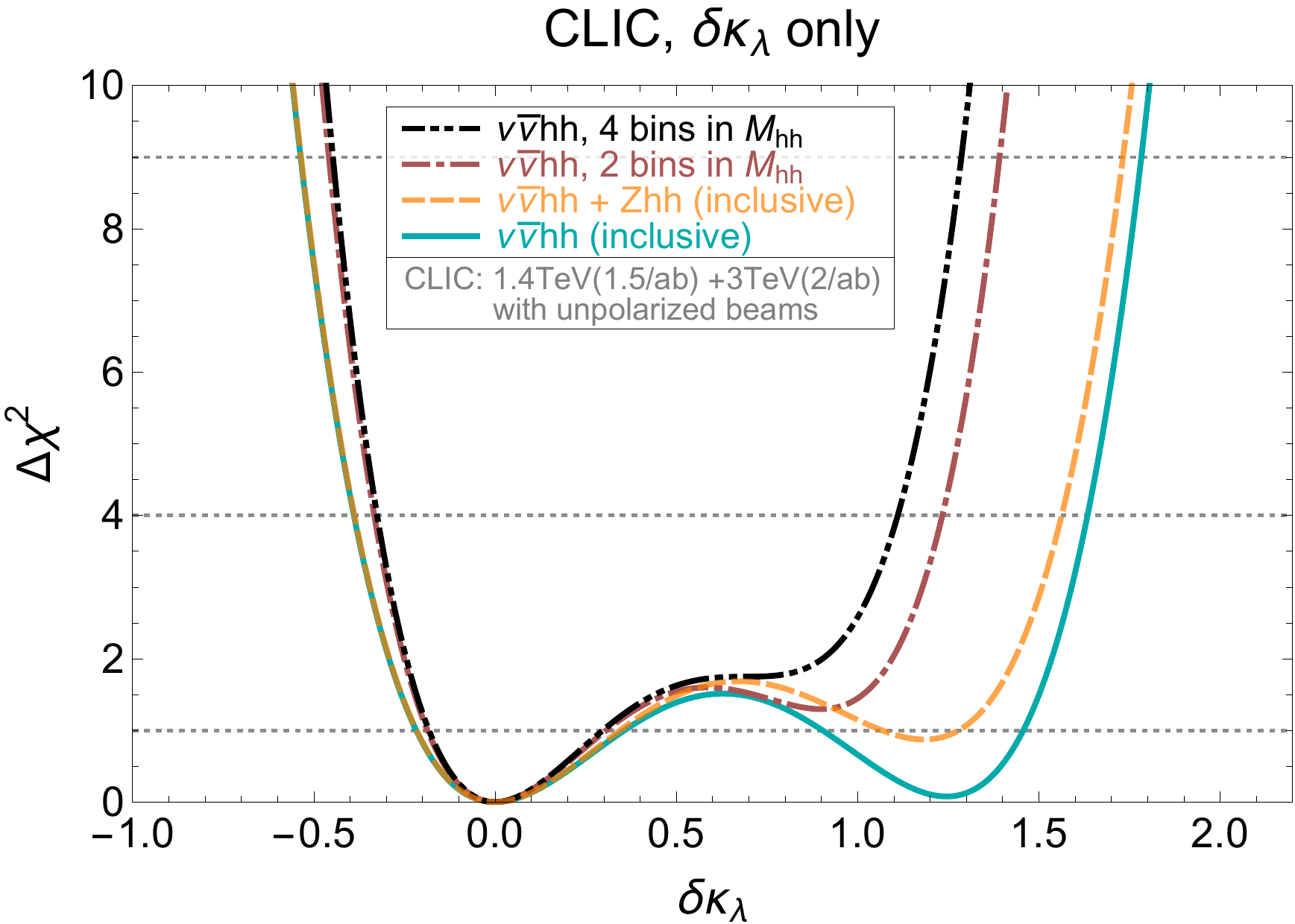}
\caption{Chi-square for the exclusive fit of $\dkl$ for various combinations of Higgs pair production measurements at the ILC (left) and CLIC (right).
}
\label{fig:lcchis1}
\end{figure}

From \autoref{fig:hhsen1}, one can see that the interference between diagrams with and without a trilinear Higgs vertex
has opposite sign in double Higgsstrahlung and $WW$-fusion.
These two processes are thus more sensitive to positive and negative values of $\dkl$ respectively. A combination of double Higgsstrahlung and $WW$-fusion measurements could hence be used to maximize the precision for both positive and negative values of $\dkl$.
Such a scenario could be achieved at the ILC through the combination of a $500\,$GeV and a $1\,$TeV run.
The impact of such combination can be clearly seen from the plot in the left panel of \autoref{fig:lcchis1}.

Being quadratic functions of $\dkl$, inclusive cross sections (for each process and collider energy)
can match the SM ones not only for $\dkl = 0$, but also for an additional value of the trilinear Higgs self-coupling,
resulting in a second minimum in the $\Delta \chi^2$.
In $WW$-fusion, the SM cross section is also obtained for $\dkl\simeq 1.08$, $1.16$ and $1.30$ at center-of-mass energies of $1$, $1.4$ and $3\,$TeV, respectively. Whereas, for double Higgsstrahlung at $500\,$GeV, the SM cross section is recovered at $\dkl \simeq -5.8$. This latter solution poses no practical problem for ILC since it can be excluded by HL-LHC measurements. Alternatively, it can be constrained by Higgs pair production through $WW$-fusion at $1\,$TeV, as well as through the indirect sensitivity of single Higgs measurements.

For CLIC, the secondary solutions
at $\dkl \simeq 1$ are more problematic. They can be constrained neither by HL-LHC data, nor by single Higgs measurements which are mostly efficient close to the threshold of the single Higgsstrahlung production.
A more promising possibility is to exploit double Higgsstrahlung rate measurements. At center-of-mass energies above $1\,$TeV, however, they only provide weak handles on $\dkl$. The $\eezhh$ cross section becomes relatively small, being only $0.08\,$fb at $1.4\,$TeV with unpolarized beams. Moreover, the sensitivity to the trilinear Higgs self-coupling decreases with energy, as shown in \autoref{fig:hhsen1}. Since the experimental collaborations did not provide an estimate for the CLIC precision achievable on the SM $e^+ e^- \rightarrow Zhh$ rate, we estimate it by naively rescaling the ILC $500\,$GeV projections by the total cross section at CLIC. We find that adding this information to inclusive $\eevvhh$ rates measurements only excludes the second minimum to the $1\sigma$ level (dashed orange line in the right panel of \autoref{fig:lcchis1}).

\begin{figure}[t]
\centering
\includegraphics[width=.47\textwidth]{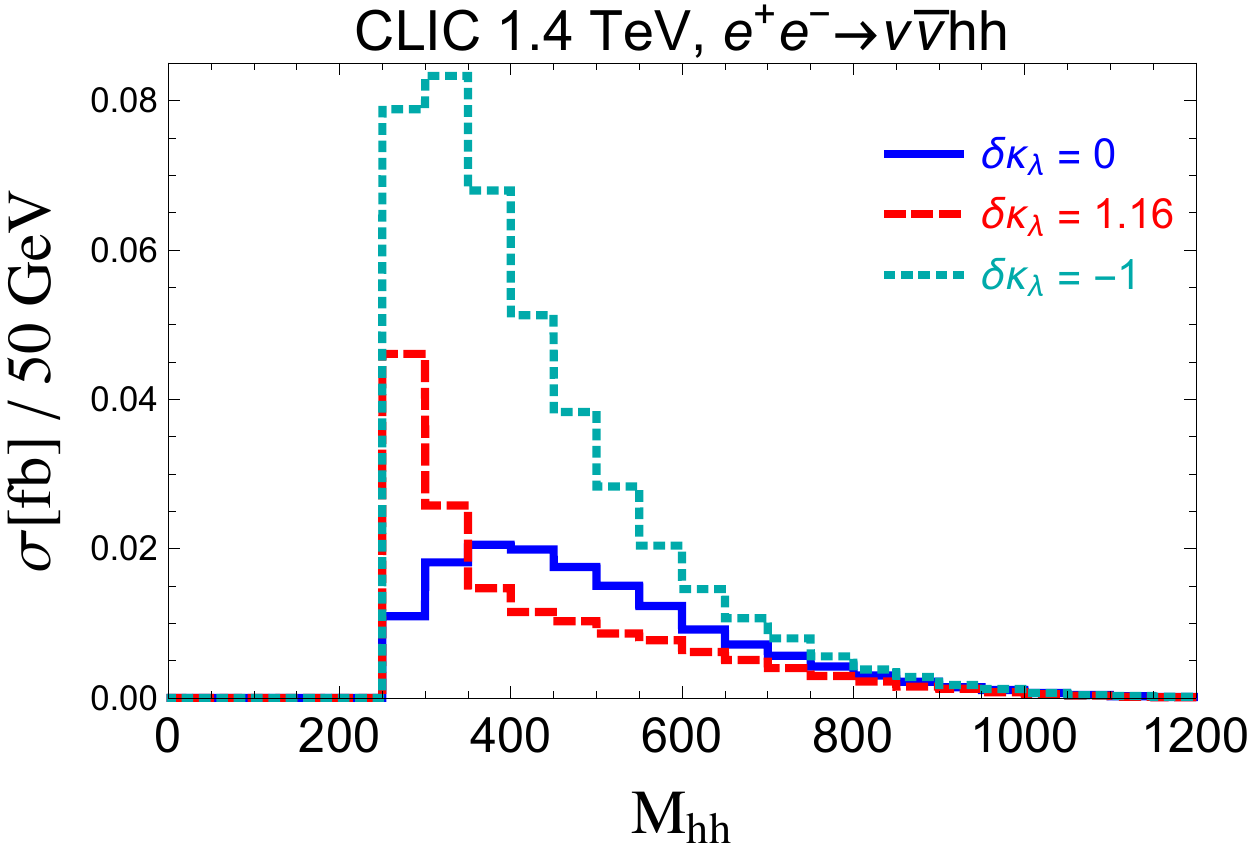}
\hfill
\includegraphics[width=.48\textwidth]{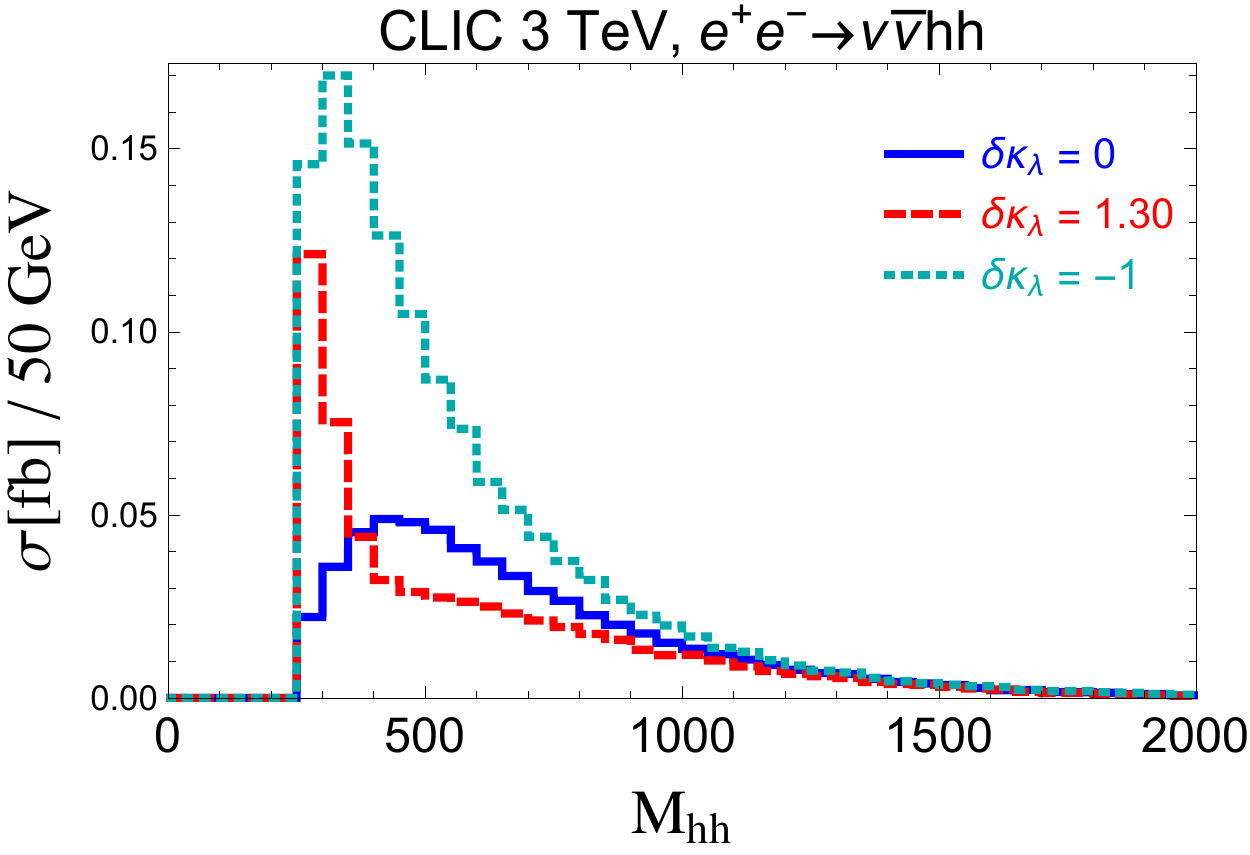}
\caption{%
Invariant mass distribution of the Higgs pair in $\eevvhh$ at $1.4\,$TeV (left) and $3\,$TeV (right). The solid blue curves are obtained in the SM ($\dkl=0$). The red dashed curves are obtained with the other value of $\dkl$ which leads to a cross section equal to the SM one. The cyan dotted curves are obtained for vanishing Higgs self-coupling ($\dkl=-1$).
}
\label{fig:clicmhh1}
\end{figure}

In addition, we consider the possibility of performing a differential analysis of double Higgs production through $WW$-fusion, studying whether a fit of the Higgs pair invariant mass distribution $M_{hh}$ can be sufficient to further exclude the $\dkl \simeq 1$ points.
The $M_{hh}$ distribution shows a good sensitivity to the Higgs trilinear, which mainly
affects the shape of the distribution close to the kinematic threshold.
This can be observed in \autoref{fig:clicmhh1}, obtained at the parton level with \texttt{MadGraph5}~\cite{Alwall:2014hca} (with \texttt{FeynRules}~\cite{Alloul:2013bka} and the \texttt{BSMC Characterisation} model~\cite{bsmc}) for $1.4$ and $3\,$TeV center-of-mass energies. The solid blue curves correspond to the SM point $\dkl = 0$.
The dashed red curves are obtained for the other value of $\dkl$ at which the $\vvhh$ coincides with the SM value ($\dkl=1.16$ for $1.4\,$TeV and $\dkl=1.30$ for $3\,$TeV).
The dotted cyan distributions are obtained for vanishing trilinear Higgs self-coupling ($\dkl=-1$).

We estimate the impact of a differential analysis of the $\vvhh$ channel by performing a simple
fit of the $M_{hh}$ invariant mass distribution. We consider either two or four bins,
whose ranges are listed in \autoref{tab:clicbin}.
For simplicity, we work at parton level and assume a universal signal over background ratio across all bins.
The right panel of \autoref{fig:lcchis1}
summarizes the result of the fits.
It shows that a differential analysis can be useful in enhancing the precision on $\dkl$. In particular,
it allows us to exclude the second fit solution $\dkl \simeq 1.3$ at the $68\%$ CL, and to reduce significantly the
$95\%$ CL bounds for positive deviations in the Higgs self-coupling. For instance, the $4$-bin fit restricts $\dkl$
to the range $[-0.18, 0.30]$ at $68\%$ CL and $[-0.33, 1.11]$ at $95\%$ CL.

\begin{table}[tb]
\centering
\begin{tabular}{c||c|c||c|c|c|c}
\rule[-.5em]{0pt}{1.em}& \multicolumn{2}{|c||}{2 bin boundaries [GeV]}  & \multicolumn{4}{c}{4 bin boundaries [GeV]}  \\ \hline\hline
\rule[-.5em]{0pt}{1.6em}1.4\,TeV   & \hspace{.25em} 250-400 \hspace{.25em}  & 400-1400   & 250-350  &  350-500   & 500-600   &  600-1400     \\ \hline
\rule[-.5em]{0pt}{1.6em}3\,TeV      & 250-500  &  500-3000  & 250-450  &  450-650  & 650-900   &  900-3000      
\end{tabular}
\caption{Definitions of the bins used in the Higgs-pair invariant mass distribution of $\eevvhh$ at $1.4\,$TeV and $3\,$TeV.}
\label{tab:clicbin}
\end{table}

\subsection{Global analysis}
\label{sec:global2h}

It is important to verify whether the results discussed in \autoref{sec:exp2h}, obtained assuming new physics affects only the triple Higgs coupling, are robust in a global framework once all other EFT parameters are taken into consideration.
We therefore perform a global analysis at ILC and CLIC including measurements of both double-Higgs (Higgsstrahlung and $WW$-fusion) and single-Higgs processes ($\vvh$, $Zh$, $t\overline t h$ and $e^+ e^- h$) in addition to diboson production.

We adopt the following benchmark scenarios chosen by the experimental collaborations for Higgs measurement estimates:
\begin{itemize}
\item \textbf{ILC}: we follow the scenario in Ref.~\cite{Barklow:2015tja}, assuming ILC can collect $2\inab$ at $250\,$GeV, $200\infb$ at $350\,$GeV and $4\inab$ at $500\,$GeV, equally shared between the $P(e^-,e^+)=(\pm0.8,\mp0.3)$ beam polarizations. We also consider the possibility of an additional run at $1\,$TeV gathering $2\inab$ with one single $P(e^-,e^+)=(-0.8,+0.2)$ beam polarization.

\item\textbf{CLIC}: we follow Ref.~\cite{Abramowicz:2016zbo} and assume $500\infb$ at $350\,$GeV, $1.5\inab$ at $1.4\,$TeV and $2\inab$ at $3\,$TeV can be collected with unpolarized beams. It should be noted that a left-handed beam polarization could increase the $\vvhh$ cross section and somewhat improve the reach on $\dkl$.

\end{itemize}

For the global fit, we follow the procedure and assumptions adopted for the single Higgs processes fit at
low-energy colliders.
We also include the one-loop dependence on $\dkl$ in single Higgs production and decay processes, as done in \autoref{sec:single}. Such effects are also included in the top-Higgs associated production $\eetth$ and in $ZZ$-fusion $\eeeeh$, although they have a negligible impact.
On the other hand, only the tree-level Higgs self-coupling dependence is considered in Higgs pair production processes, since one-loop corrections are numerically insignificant.
As already stressed, the quadratic dependence on $\dkl$ in Higgs pair production processes cannot be neglected. In this case, cross terms between $\dkl$ and other EFT parameters are also accounted for. The linear approximation is adopted elsewhere.
The estimates for the precision of the SM Higgs pair production cross section are taken from Refs.~\cite{Duerig:2016dvi, Abramowicz:2016zbo, ILC1TeV} already discussed in the previous section.

\begin{figure}[t]
\centering
\includegraphics[width=.48\textwidth]{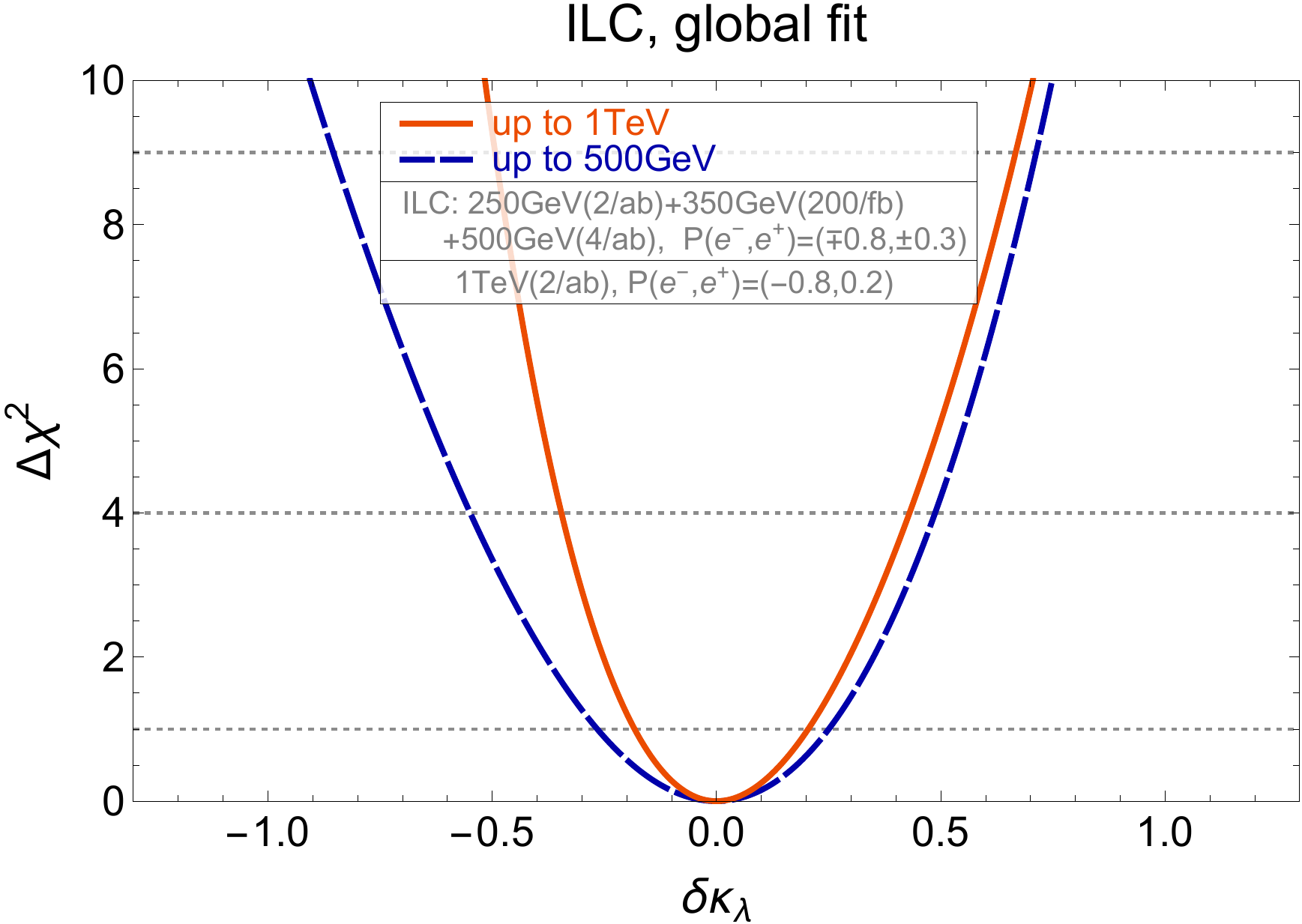}
\hfill
\includegraphics[width=.48\textwidth]{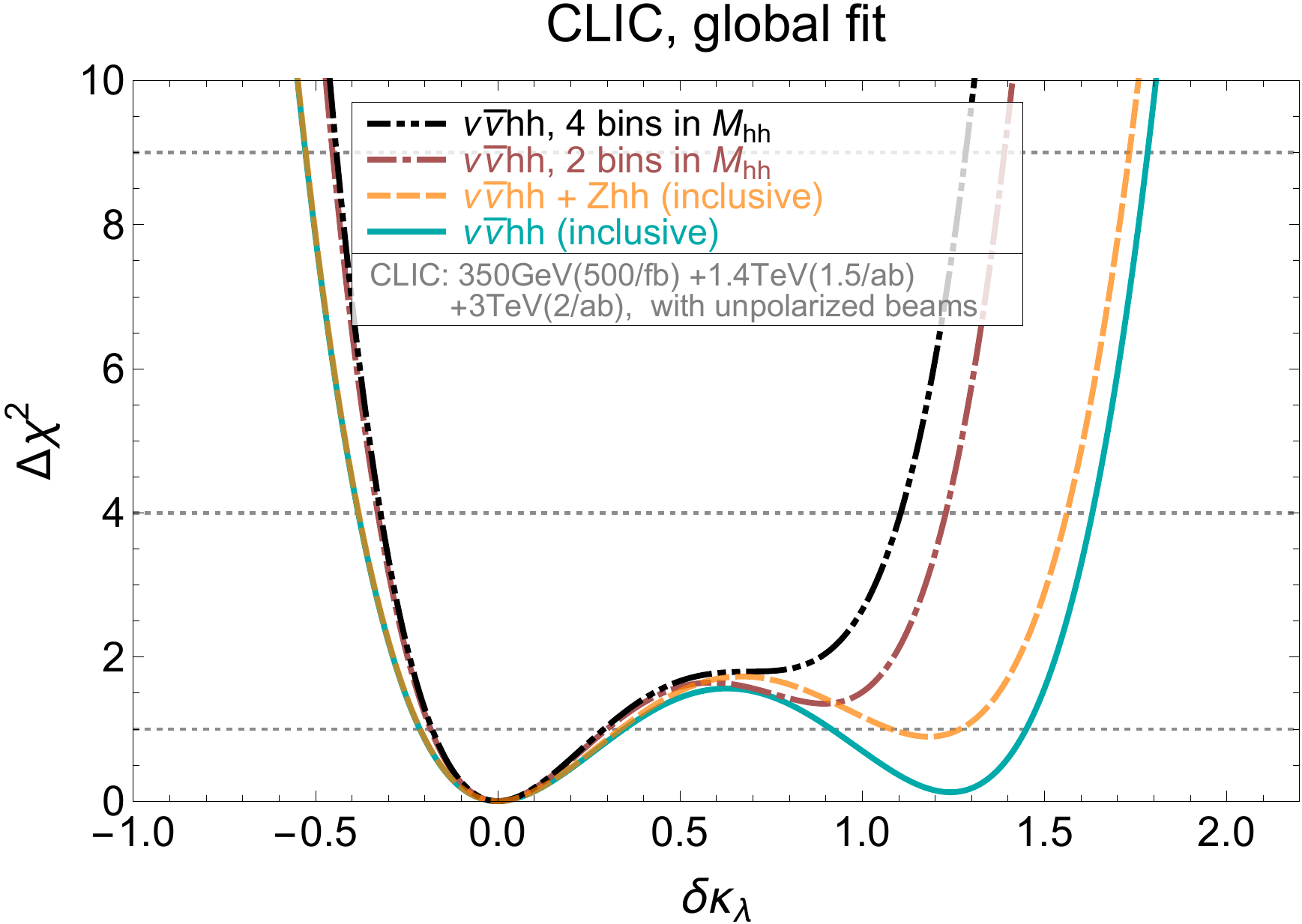}
\caption{Chi-square as a function of $\dkl$ for the high-energy ILC (left) and CLIC (right) benchmarks. The results are obtained through a global analysis, profiling over all other EFT parameters.
}
\label{fig:lcchis2}
\end{figure}

The results of the global fit for the ILC and CLIC benchmark scenarios are shown in \autoref{fig:lcchis2}.
The $68\%$ and $95\%$ CL intervals are also listed in \autoref{tab:lc2}.
It is interesting to compare these results with the ones obtained through the exclusive fit on $\dkl$ discussed in
\autoref{sec:exp2h} (see \autoref{fig:lcchis1}).
The $\chi^2$ curves for ILC (up to $500\,$GeV or $1\,$TeV) and CLIC (no binning, 2 bins and 4 bins in $M_{hh}$)
show very mild differences in the global fit with respect to the exclusive one. This demonstrates that
the additional EFT parameters are sufficiently well constrained by single Higgs measurements and therefore have a marginal impact
on the global fit.
We also analyzed the impact of combining ILC and CLIC measurements with HL-LHC ones. The precision achievable at the LHC is significantly poorer than the one expected at high-energy lepton colliders, so that the latter dominate the
overall fit and only a mild improvement is obtained by combination.

\begin{table}[t]
\centering
\begin{tabular}{c|c|c}
& 68 \%CL & 95\%CL\\
\hline
\hline
\rule[-.5em]{0pt}{1.6em}ILC up to $500\,$GeV   &  $[-0.27,~0.25]$    &  $[-0.55,~0.49]$    \\ \hline
\rule[-.5em]{0pt}{1.6em}ILC up to $1\,$TeV   &  $[-0.18,~0.20]$    &  $[-0.35,~0.43]$    \\ \hline\hline
\rule[-.5em]{0pt}{1.6em}CLIC  &  $[-0.22,~0.36] \cup [0.91,~1.45]$     &  $[-0.39,~1.63]$    \\ \hline
\rule[-.5em]{0pt}{1.6em} $+Zhh$   &  $[-0.22,~0.35] \cup [1.07,~1.27]$     &  $[-0.39,~1.56]$    \\ \hline
\rule[-.5em]{0pt}{1.6em} 2 bins in $\vvhh$   &  $[-0.19,~0.31]$   &  $[-0.33,~1.23]$    \\ \hline
\rule[-.5em]{0pt}{1.6em} 4 bins in $\vvhh$   &  $[-0.18,~0.30]$   &  $[-0.33,~1.11]$

\end{tabular}
\caption{
Precision on the determination of $\dkl$ obtained through a global fit including pair- and single-Higgs production
channels for several benchmark scenarios at ILC and CLIC.
}
\label{tab:lc2}
\end{table}

\begin{figure}[t]
\centering
\includegraphics[width=.48\textwidth]{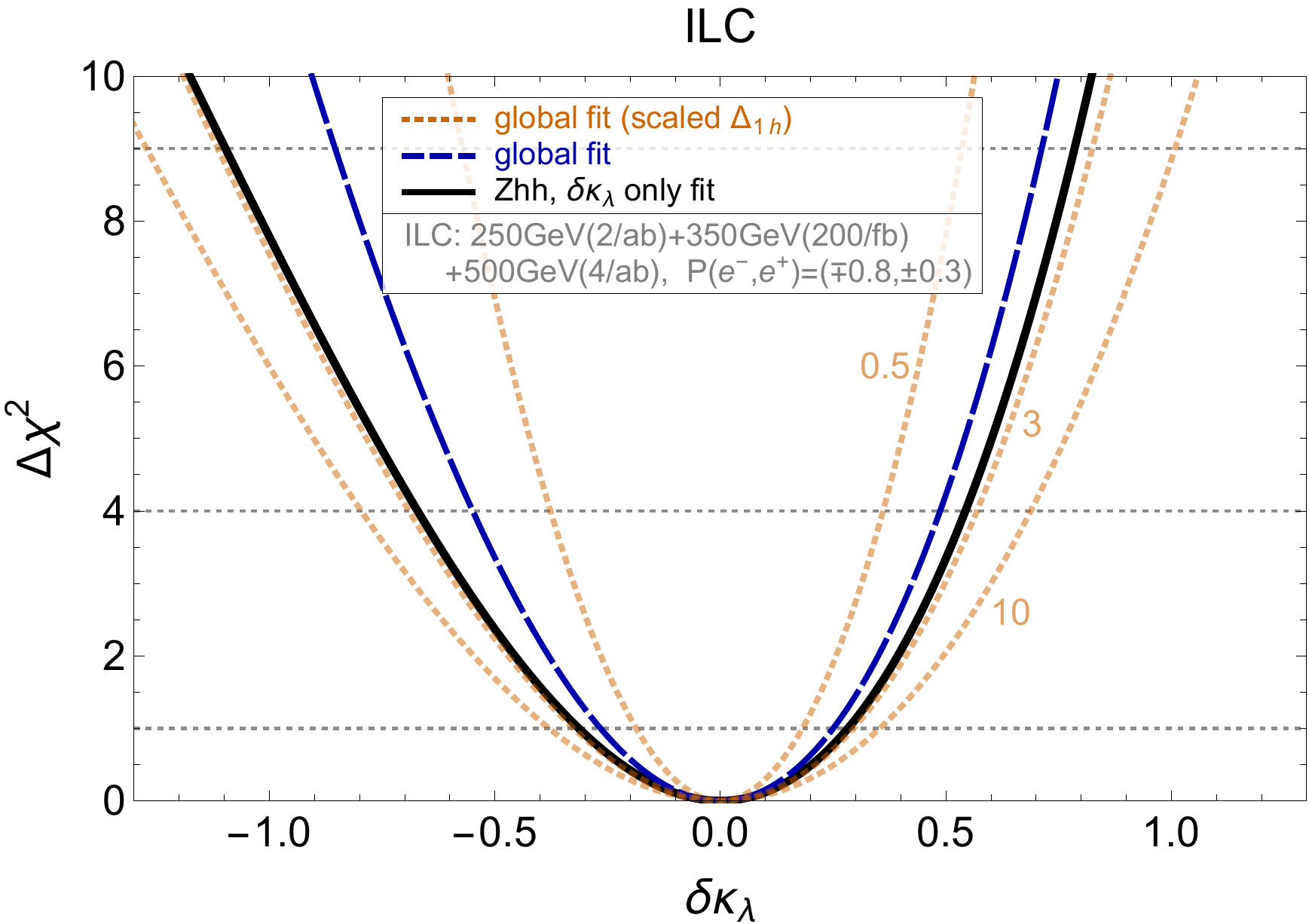}   \hfill
\includegraphics[width=.48\textwidth]{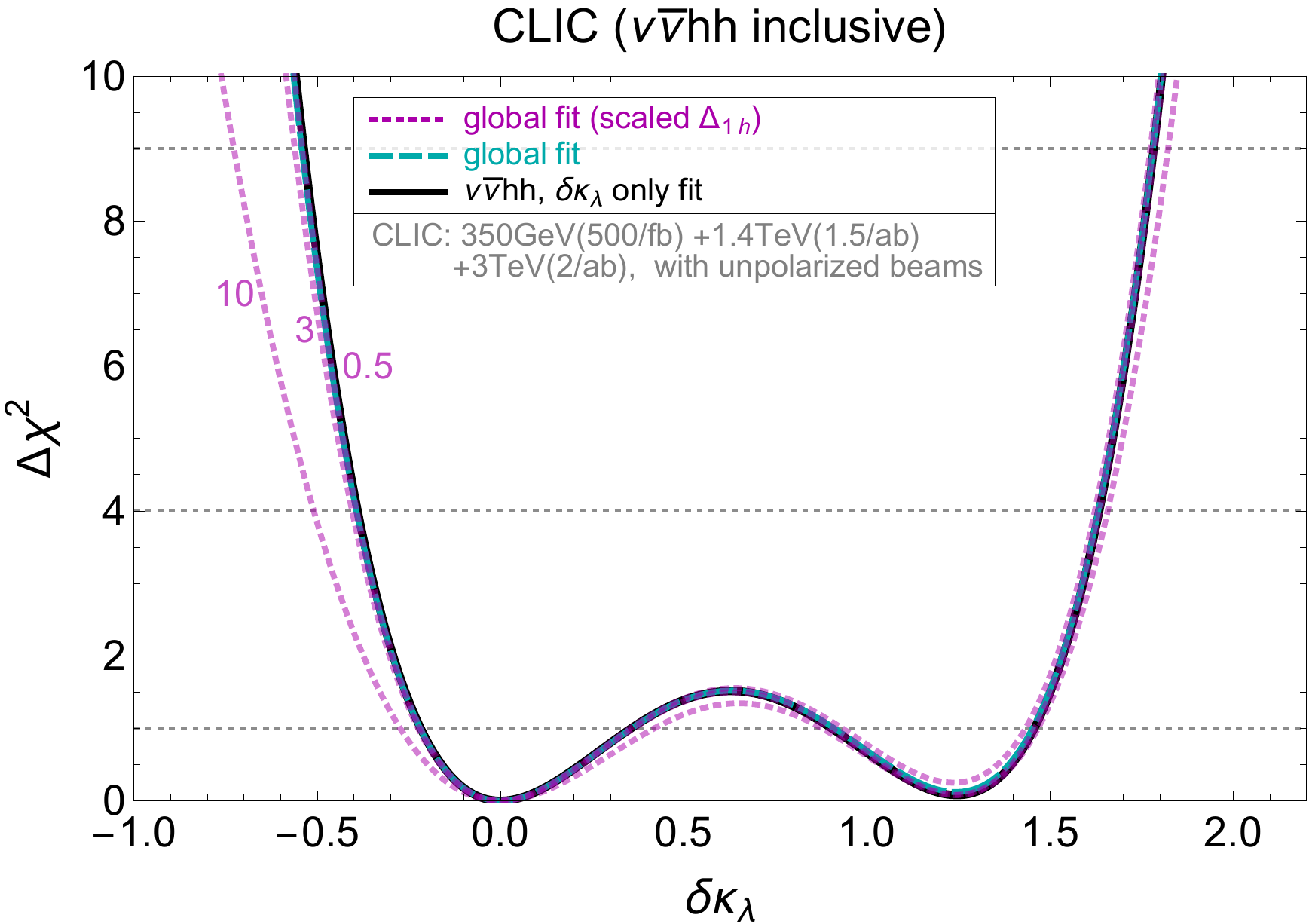} 
\caption{ {\bf Left:} Chi-square profiled over all EFT parameters but $\dkl$ for ILC (up to $500\,$GeV).  {\bf Right:} The same for CLIC (no binning in $M_{hh}$).  Three scenarios are shown.  The solid black curves correspond to the $\dkl$ only fit from the double-Higgs measurements.  The dashed blue/cyan curves correspond to the global fits  in \autoref{fig:lcchis2}.
The additional dashed curves are obtained by rescaling the uncertainties of single Higgs measurements (including $\eeww$) by an overall factor.  For example, $\Delta_{1h}\times 10$ denotes that the uncertainties of the single Higgs and diboson measurements are multiplied (worsened) by a factor $10$.
}
\label{fig:lcchis3}
\end{figure}

\medskip

We saw that allowing for other EFT deformations beside $\dkl$ does not worsen the global fit significantly. This result, however, was by no means guaranteed. To stress this point, we display in \autoref{fig:lcchis3} the profiled $\chi^2$ obtained by artificially rescaling the precision in single Higgs measurements. The ILC (up to 500\,GeV, left panel) and CLIC (no binning in $M_{hh}$, right panel) benchmarks are used as examples.
For each collider, we show the results of the exclusive $\dkl$ analysis of the Higgs pair production measurements (solid black curve) and of the global analysis (dashed blue/cyan). The additional dashed curves correspond to global fits in which
the precision in single Higgs and diboson measurements is rescaled by factors ranging from $0.5$ to $10$.
It can be seen that the global fit is sizably affected by such a rescaling, in particular the fit precision is significantly degraded if single Higgs measurements become worse.
This result shows that a comprehensive global analysis of the single Higgs measurements is crucial for obtaining robust constraints on $\dkl$.
Notice moreover that an improved precision on single Higgs measurements could have a positive impact
on the determination of the Higgs self coupling at the ILC.

\medskip

The impact of the uncertainty on the EFT parameters measurements on the extraction of the Higgs self-coupling from
Higgs pair production was also recently investigated in Ref.~\cite{Barklow:2017awn}. It focused mainly on
Higgs pair production through double Higgsstrahlung at ILC $500\,$GeV and on single-Higgs production in lower-energy
runs, taking into account the uncertainties on SM parameters and electroweak precision observables.
Loop-level contributions to single-Higgs processes coming from a modified Higgs self-coupling were not included in the fit,
and the linear approximation was used to obtain the numerical results.
The final fit takes into account runs at $250$ and $500\,$GeV, with $2$ and $4\inab$ respectively equally shared between $P(e^-,e^+)=(\mp0.8,\pm0.3)$ beam polarizations. The estimated precision on the measurement of $\dkl$ is
$30\%$, which is in good agreement with the constraints we obtained in our ILC benchmark scenario.

%%%%%%%%%%%%%%%%%%%%%%%%%%%%%%%%%%%%%%%%%%%%%%%%%%%%%%%%%%%%%%%%%%
\section{Summary and conclusions}
\label{sec:sum}

In this paper, we analyzed the precision reach on the determination of the Higgs trilinear self-coupling at future
lepton colliders.  We covered a comprehensive set of scenarios including low-energy and high-energy machines.
The former can only access the Higgs self-interaction indirectly through NLO corrections to single Higgs processes.
High-energy colliders can instead test deviations in the Higgs trilinear coupling directly, through the measurement of
Higgs pair production, in particular double Higgsstrahlung and $WW$-fusion.

\begin{figure}[t]
\centering
\includegraphics[width=0.95\textwidth]{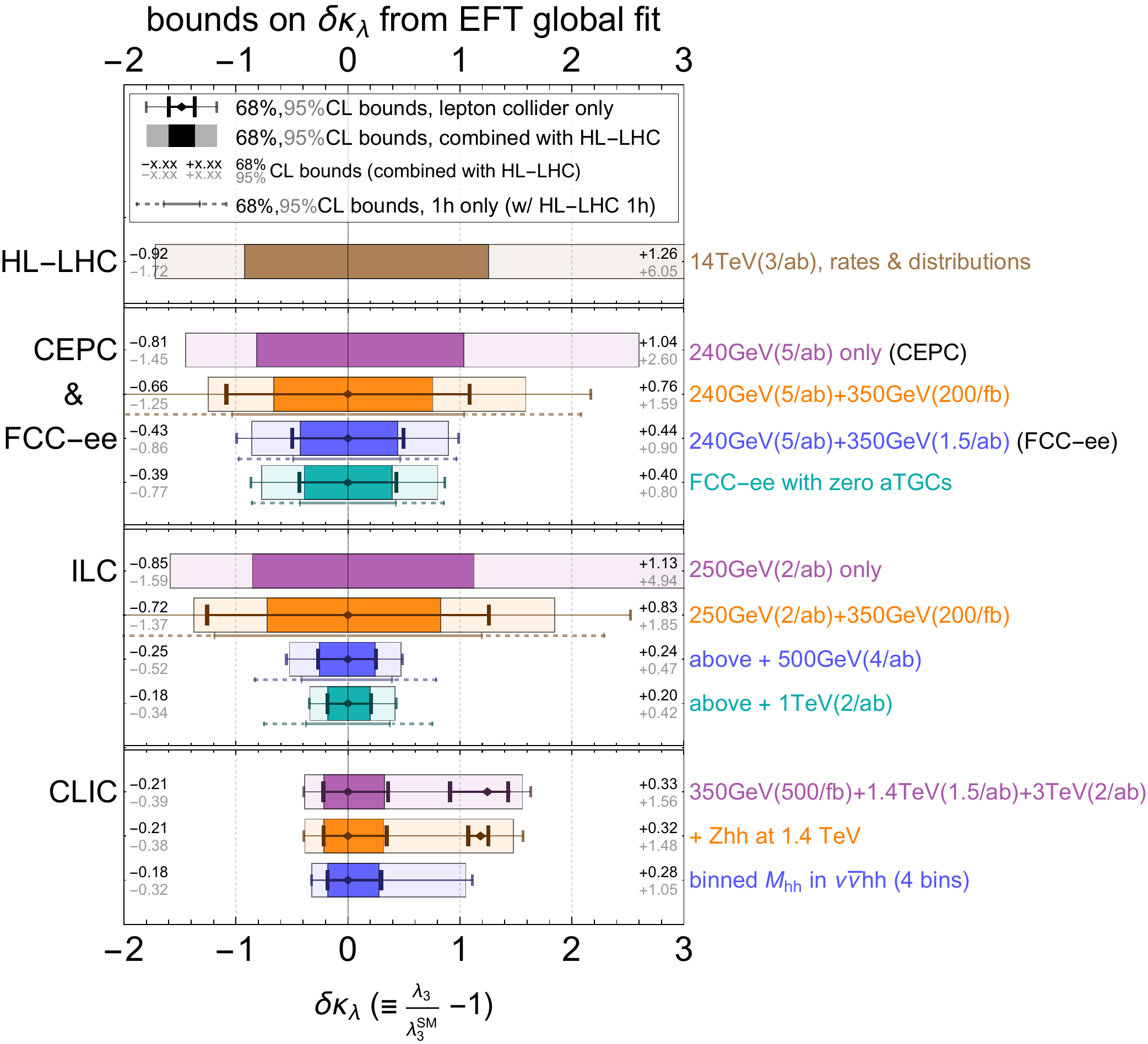}
\caption{A summary of the bounds on $\dkl$ from global fits for various future collider scenarios.  For the ``$1h$ only'' scenario, only single Higgs measurements at lepton colliders are included.
}
\label{fig:sum1}
\end{figure}

We performed a global analysis, simultaneously taking into account corrections to the Higgs self-coupling and deviations
in EFT parameters affecting Higgs interactions with other SM particles. The results of the analysis are summarized in \autoref{fig:sum1} for the various benchmark scenarios considered. For each scenario, three sets of bounds are shown. Thin lines with vertical ends show the precision expected from measurements at lepton colliders only. The superimposed thick bars combine them with HL-LHC measurements. Finally, the thin solid and dotted lines are obtained by combining single Higgs measurements only at lepton colliders ($1h$) with the HL-LHC bounds.
As discussed in the main text, unpolarized beams are assumed for the CEPC, FCC-ee and CLIC. For the ILC runs up to $500\,$GeV, an equal share of the luminosity at the two $P(e^-, e^+) = (\pm0.8, \mp0.3)$ beam polarizations is assumed, whereas a single polarization $P(e^-, e^+) = (-0.8 , +0.2)$ is adopted at $1\,$TeV.

We found that a global analysis is essential to derive robust bounds on $\dkl$.
This is the case, in particular, if only low-energy lepton machines, such as CEPC or FCC-ee, are available.
In this scenario, the Higgs self-coupling can be determined with good accuracy, around $40\%$ at the $68\%$ CL,
by exploiting single Higgs measurements in the $\vvh$ and $Zh$ channels as well as diboson production.
In order to achieve this accuracy, it is essential to combine runs at different center-of-mass energy, for instance
at $240\,$GeV and at $350\,$GeV, both with luminosities in the few attobarns range. Measurements at a single energy, in fact, leave a nearly flat direction unresolved in the global fit and lead to a very poor determination of $\dkl$. Runs at two different energies can instead significantly reduce the flat direction by constraining with better accuracy on the other EFT parameters.

The high-energy linear colliders making direct measurements of the triple Higgs self-coupling
through pair production still provide the best constraints. Double Higgsstrahlung and $WW$-fusion yield complementary
information, being more sensitive to positive and negative deviations in the Higgs self-coupling respectively.
It is interesting to notice that the dependence of these two processes on $\dkl$ is stronger at lower center-of-mass
energy, as shown in \autoref{fig:hhsen1}, so that ILC runs at $500\,$ GeV and $1\,$TeV energy maximize the overall precision allowing for a determination
of the trilinear Higgs self-coupling with a $20\%$ uncertainty approximately, at the $68\%$~CL.

High-energy measurements alone, such as the ones available with the $1.4$ and $3\,$TeV CLIC runs, can only rely
on $\vvhh$ production and have limited sensitivity to positive deviations in $\dkl$. In this case, a second minimum in the
global fit is present for $\dkl \sim 1$. The additional minimum can be excluded by performing a differential analysis
exploiting the Higgs pair invariant mass distribution, whose threshold behavior is strongly sensitive to deviations in
the Higgs self-coupling. A differential analysis can provide an order-$20\%$ determination of $\dkl$ at $68\%$ CL,
however at $95\%$ CL values $\dkl \simeq 1$ would still be allowed.

It is interesting to compare the above results with the ones achievable at the HL-LHC and at
possible future hadron colliders. The HL-LHC is expected to be sensitive only to deviations of ${\cal O}(1)$ in the
Higgs self-coupling. As one can see from \autoref{fig:sum1}, this precision is comparable to (or better than)
the one achievable at low-energy lepton colliders with low integrated luminosity at $350\,$GeV runs.
This is the case for our circular collider benchmarks with $200\infb$ integrated luminosity at $350\,$GeV, as well as
for the low-energy runs of the ILC. In these scenarios the HL-LHC data will still play a major role in
the determination of $\dkl$, while lepton colliders always help constraining large positive $\dkl$ that the HL-LHC fails to exclude beyond the one-sigma level.
On the other hand, with $1\inab$ of luminosity collected at $350\,$GeV, the lepton collider data starts dominating the combination.

The situation is instead different at high-energy hadron colliders which can benefit from a sizable cross section in
double Higgs production through gluon fusion. A $pp$ collider with $100\,$TeV center-of-mass energy is expected to determine $\dkl$ with a precision of order $5\%$~\cite{Contino:2016spe}, thus providing a better accuracy than
lepton machines. Intermediate-energy hadron machines, such as a high-energy LHC at $27-33\,$TeV could instead
provide a precision comparable to that of high-energy lepton colliders. A rough estimate of the $\dkl$ determination
at a $33\,$TeV $pp$ collider gives a $\sim 30\%$ precision at $68\%$ CL for an integrated luminosity of $10\inab$.

To conclude the discussion, let us come back to our assumption of perfectly well measured electroweak precision observables.
It seems fully justified if low-energy runs at the $Z$-pole are performed. This could for instance be the case at the ILC, CEPC, and FCC-ee which could respectively produce $10^9$, $10^{10}$, and $10^{12}$ $Z$ bosons.
A $Z$-pole run for these machines can provide significant improvements with respect to LEP measurements ($2\cdot 10^7$ $Z$ bosons), making electroweak precision observables basically irrelevant for the extraction of the Higgs trilinear self-coupling.

Without a new $Z$-pole run, evaluating the impact of a limited accuracy on electroweak precision observables might be less straightforward. An analysis of such scenario for the ILC collider has been recently presented in Ref.~\cite{Barklow:2017awn}. This work explicitly includes present constraints on $m_Z$, the $A_\ell$ asymmetry at the $Z$-pole, $\Gamma_{Z\to ll}$, $\Gamma_Z$, $\Gamma_W$ and forecasts for improved $m_W$, $m_H$, and $\Gamma_W$ measurements, assuming no new run at the $Z$-pole. In that scenario, it is argued that Higgs measurements can be used to improve the constraints on the electroweak parameters. The achievable precision is sufficient to ensure that
electroweak precision observables do not significantly affect the determination of $\dkl$.

The precision necessary to decouple electroweak and Higgs parameters determinations in other benchmark scenarios might deserve further exploration. We think that electroweak precision measurements will have a negligible impact on trilinear Higgs self-coupling determination at high-energy machines where Higgs pair production is accessible. This conclusion is supported by the results of \autoref{sec:double} showing that the determination of $\dkl$ is only mildly affected by the other EFT parameters, once a wide-enough set of single Higgs measurements is considered. The situation for low-energy colliders, in which the Higgs self-coupling can be accessed only indirectly through single Higgs processes, is instead less clear. As we saw in \autoref{sec:single}, the precision on $\dkl$ obtained through a global fit is significantly lower than the one estimated through an exclusive analysis.
Consequently, the precision of the single-Higgs and triple-gauge coupling extractions has a relevant impact on the fit. In principle, electroweak precision parameters could affect the bounds on single Higgs couplings and thus indirectly degrade the $\dkl$ constraint. This aspect might be worth a more careful investigation, which is however beyond the scope of the present work.

\section*{Acknowledgments}

We thank T.~Barklow, P.~Janot, M.~McCullough, F.~Maltoni, M.~Peskin and L.-T.~Wang for helpful discussions.  We thank W.~H.~Chiu, S.~C.~Leung, T.~Liu, K.-F.~Lyu and L.-T.~Wang for coordinated publication of their related work.

JG is supported by an International Postdoctoral Exchange Fellowship Program between the Office of the National Administrative Committee of Postdoctoral Researchers of China (ONACPR) and DESY.
C.G. is supported  by the Helmholtz Association. M.R. is supported by la Caixa, Severo Ochoa grant program. G.P., M.R. and T.V. are supported by the Spanish Ministry MEC under grants FPA2015-64041-C2-1-P, FPA2014-55613-P and FPA2011-25948, by the Generalitat de Catalunya grant 2014-SGR-1450 and by the Severo Ochoa excellence program of MINECO (grant SO-2012-0234). G.P is also supported by the European Commission through the Marie Curie Career Integration Grant 631962 via the DESY-IFAE cooperation exchange. We thank the Collaborative Research Center SFB676 of the Deutsche Forschungsgemeinschaft (DFG), ``Particles, Strings and the Early Universe'', for support.
This manuscript has been authored by Fermi Research Alliance, LLC under Contract No. DE-AC02-07CH11359 with the U.S. Department of Energy, Office of Science, Office of High Energy Physics. The United States Government retains and the publisher, by accepting the article for publication, acknowledges that the United States Government retains a non-exclusive, paid-up, irrevocable, world-wide license to publish or reproduce the published form of this manuscript, or allow others to do so, for United States Government purposes.

%%%%%%%%%%%%%%%%%%%%%%%%%%%%%%%%%%%%%%%%%%%%%%%%%%%%%%%%%%%%%%%%%%

\appendix

%%%%%%%%%%%%%%%%%%%%%%%%%%%%%%%%%%%%%%%%%%%%%%%%%%%%%%%%%%%%%%%%%%

\section{One-loop corrections from \texorpdfstring{$\delta \kappa_\lambda$}{delta kappa lambda}}
\label{app:loop}

\begin{table}[h]\small
\centering
\begin{tabular}{c||c|c|c|c|c|c|c}
\rule[-.5em]{0pt}{1.1em}$C_1$ &  \multicolumn{7}{|c}{$\sqrt{s}$ [GeV]} \\  \cline{2-8}
\rule[-.5em]{0pt}{1.6em}(inclusive rates) & 240 & 250  & 350 & 500 & 1000  & 1400 & 3000   \\  \hline\hline
\rule[-.5em]{0pt}{1.6em}$\eehz$   &  0.017  & 0.015  &  0.0057  & 0.00099   &  -0.0012  &  -0.0011   & -0.00054  \\ \hline
\rule[-.5em]{0pt}{1.6em}$\eevvh\ \ ^\bigstar$   & 0.0064   & 0.0064 &  0.0062   &  0.0061   &  0.0059  &   0.0058 &  0.0057  \\ \hline 
\rule[-.5em]{0pt}{1.6em}$\eeeeh\ \ ^\bigstar$   &  0.0070   & 0.0070   &  0.0069  &   0.0067  &  0.0065  &  0.0065   & 0.0063   \\ \hline 
\rule[-.5em]{0pt}{1.6em}$\eetth$                  &               &		&		&   0.086	&  0.017  &  0.0094  &  0.0037
\end{tabular}
\caption{Values of $C_1$ for the total cross-sections of Higgs production processes. $^\bigstar\,$The numbers are for $WW$ or $ZZ$ fusion only.
}
\label{tab:c1}
\end{table}

In this appendix we collect the numerical values of the coefficients $C_1$,
defined in \autoref{eq:definec1}, which encode the corrections to single-Higgs processes due to a deformation
of the Higgs trilinear coupling.
In \autoref{tab:c1}  we report the $C_1$ coefficients for the total cross-section of the main single-Higgs production modes,
namely Higgsstrahlung, vector-boson fusion and associated production with top quarks.
Several values of the center-of-mass energy $\sqrt{s}$ are reported in the table,
corresponding to the benchmark runs of future lepton colliders considered in main text.
The calculation has been performed with the help of the public tools
\texttt{FeynArts}, \texttt{FormCalc}, \texttt{LoopTools}, and
\texttt{CUBA}~\cite{Hahn:2000kx,Hahn:1998yk,Hahn:2004fe}.

\begin{figure}\centering
\includegraphics[width=.5\textwidth]{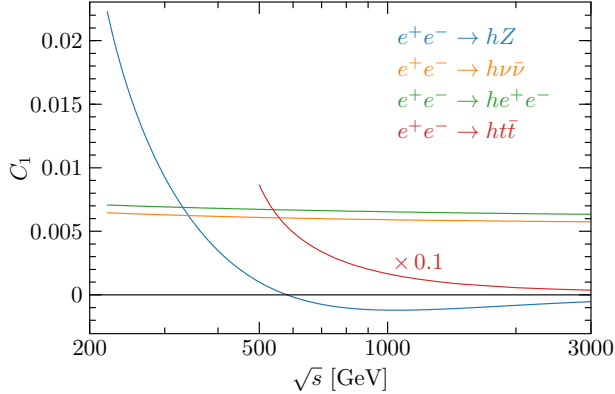}
\caption{Value of $C_1$ as a function of the center-of-mass energy $\sqrt{s}$ for the $\eehz$, $\eevvh$, $e^+ e^- \rightarrow h e^+ e^-$ and $e^+e^- \rightarrow h t \bar t$ single Higgs production processes. Notice that the result for Higgs production in association with a top-quark pair has been rescaled by a factor of $0.1$.
}
\label{fig:c1s2}
\end{figure}

Notice that the values of $C_1$ for Higgsstrahlung, $WW$-boson fusion and $ZZ$-boson fusion are independent of the beam polarization if we restrict ourselves to diagrams up to one loop, as we did in our analysis. As for $\eetth$, the Higgs self-coupling gives rise to tiny beam polarization effects. Given the small impact of the latter production mode in our analysis, we can safely neglect such effects.
The dependence of the $C_1$ coefficients on the collider energy is also shown in \autoref{fig:c1s2}.

\begin{table}[t]
\centering
\begin{tabular}{c||c|c|c|c|c}
\rule[-.5em]{0pt}{1.6em}$C_1$  & $ZZ$ & $WW$ & $\gamma\gamma$ & $gg$ & $f\bar{f}$   \\ \hline
\rule[-.5em]{0pt}{1.6em}on-shell $h$ decay & 0.0083 & 0.0073 & 0.0049 & 0.0066 & 0
\end{tabular}
\caption{Values of $C_1$ for the Higgs partial widths from Ref.~\cite{Degrassi:2016wml}.
}
\label{tab:de}
\end{table}

Besides the inclusive rates, we also checked the impact of a modified Higgs trilinear coupling
on the angular asymmetries that can be built for the $\eehz\to h\ell^+\ell^-$ case
(see Refs.~\cite{Beneke:2014sba, Craig:2015wwr}). We found that
these effects are almost negligible and have no impact on our analysis.

For completeness, we also report in \autoref{tab:de} the $C_1$ coefficients for
the Higgs partial widths~\cite{Degrassi:2016wml}.

\section{Additional results}
\label{app:results}

In this appendix, we collect some additional numerical results and plots that were not included in the main text.

\begin{figure}[t]
\centering
\includegraphics[width=.48\textwidth]{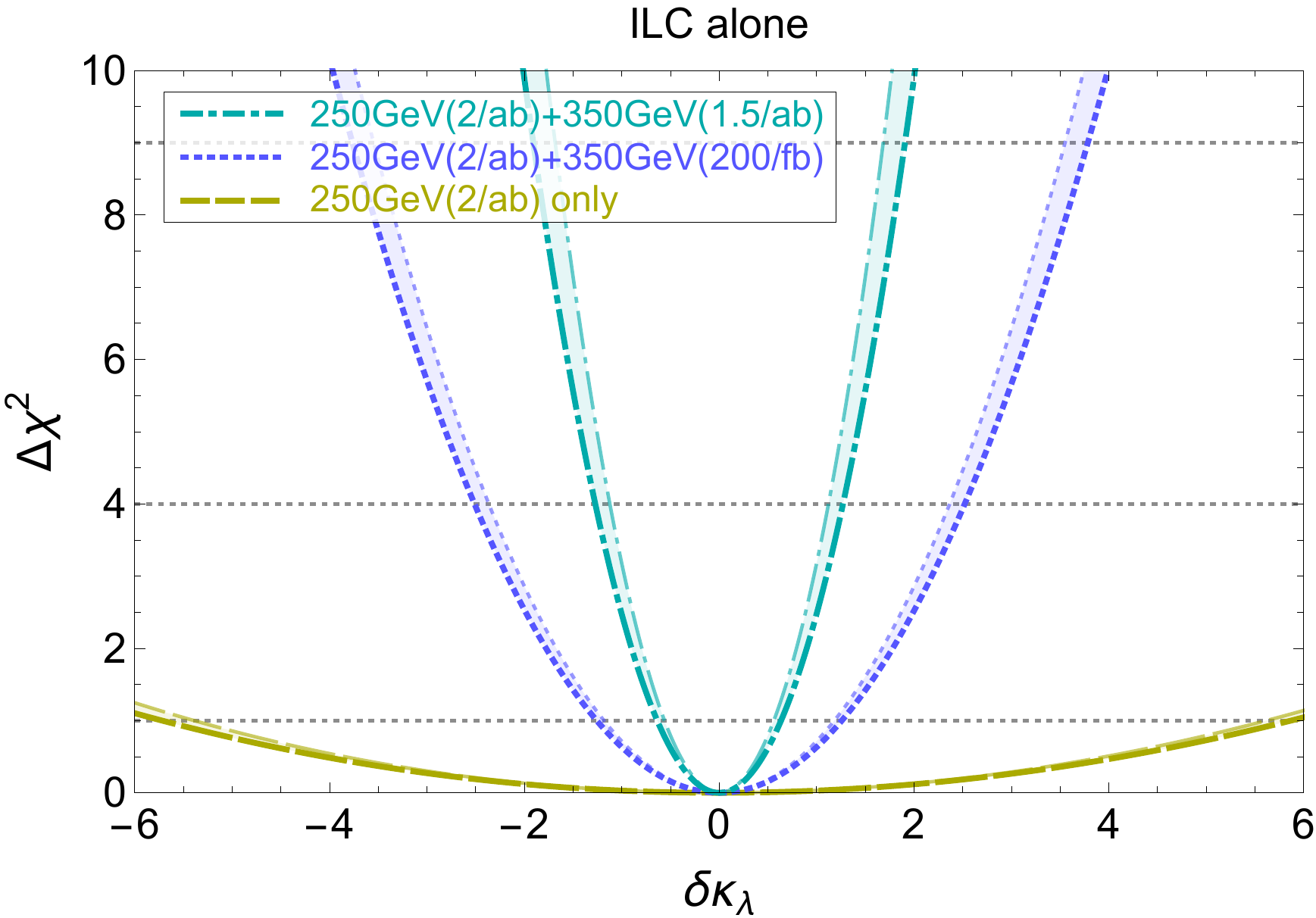}
\hfill
\includegraphics[width=.48\textwidth]{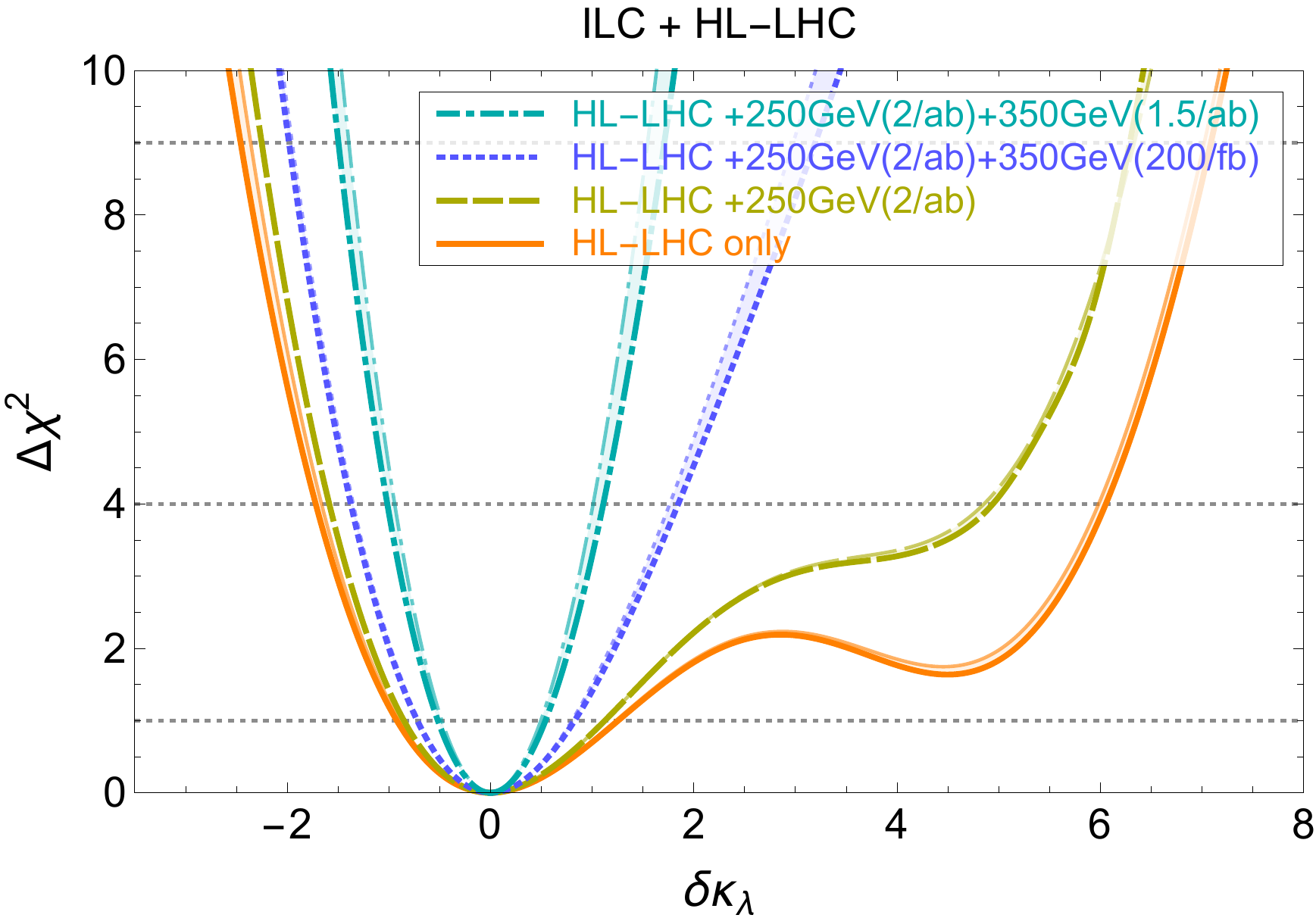}
\caption{Chi-square profiled over all EFT parameters but $\dkl$. Three run scenario are considered for ILC, with $2\inab$ at $250\,$GeV and $\{0,~200\infb,~1.5\inab\}$ at $350\,$GeV, with luminosities equally split into
$P(e^-, e^+) = (\pm0.8,\mp0.3)$ beam polarizations.
The shaded areas cover different assumptions about the precision of TGC measurements.
{\bf Left:} ILC measurements only. {\bf Right:} combination with differential single and double Higgs measurements at the HL-LHC.
}
\label{fig:npilc1}
\end{figure}

In \autoref{fig:npilc1}, we show the profiled $\Delta \chi^2$ as a function of $\dkl$ for the low-energy ILC
benchmark considered in \autoref{sec:single}, including $2\inab$ of integrated luminosity at $250\,$GeV
and either $200\infb$fb or $1.5\inab$ at $350\,$GeV with luminosities equally split into
$P(e^-, e^+) = (\pm0.8,\mp0.3)$ beam polarizations. In the left panel, we show the global fit for the ILC alone,
while in the right panel we combine these results with the differential single and double Higgs measurements at the high-luminosity LHC. The corresponding $68\%$ CL intervals are listed in \autoref{tab:resultscepc}.

In \autoref{tab:resultsilcb}, \autoref{tab:resultsilcc7030} and \autoref{tab:resultsilcc5050}, we consider three alternative
benchmark scenarios for the low-energy ILC runs. The three scenarios differ from the one considered in the main text
by different choices of beam polarizations and luminosity splitting among them. The total integrated luminosities
are the same as in the main benchmark, namely $2\inab$ at $250\,$GeV, $200\infb$fb or $1.5\inab$ at $350\,$GeV.
In \autoref{tab:resultsilcb}, we consider $P(e^-,e^+)=(\mp 0.8, \pm 0.3)$ beam polarizations with luminosity
split between them according to a $70\%/30\%$ ratio. In \autoref{tab:resultsilcc7030} and \autoref{tab:resultsilcc5050}, we consider $P(e^-,e^+)=(\mp 0.8, 0)$ beam polarizations with luminosity
split between them with a $50\%/50\%$ ratio and a $70\%/30\%$ ratio respectively.

\begin{figure}[t]\centering
\includegraphics[width=.6\textwidth]{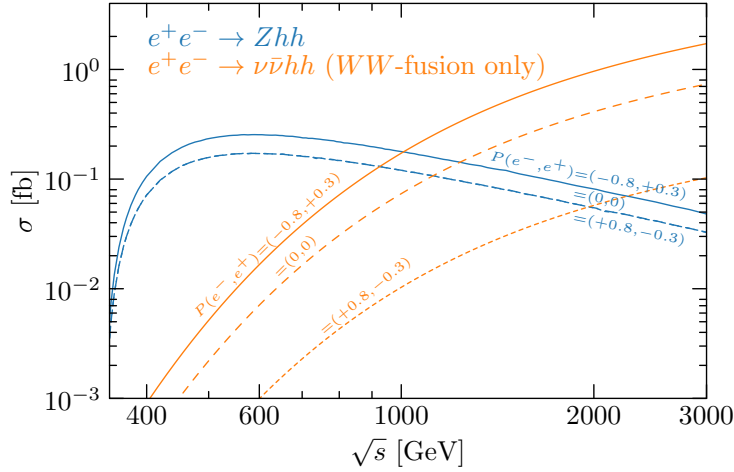}
\caption{Higgs pair production cross sections at as functions of the center-of-mass energy for different choices of the beam polarizations. The solid curves correspond to $P(e^-,e^+) = (-0.8, +0.3)$, the dotted ones to $P(e^-,e^+) = (+0.8, -0.3)$,
and the dashed one to $P(e^-,e^+) = (0, 0)$. Notice that the dashed and dotted lines for $\eezhh$ overlap with each other.
}
\label{fig:hh_xsec2}
\end{figure}

\begin{table}[t]
\centering
\begin{tabular}{@{\hspace{4pt}}c||c|c||c|c@{\hspace{4pt}}}
 $P(e^-,e^+)=(\mp 0.8, \pm 0.3)$ & \multicolumn{2}{c||}{\small ILC alone} & \multicolumn{2}{|c}{\small ILC + HL-LHC}\\
\cline{2-5}
\rule[-.4em]{0pt}{1.3em} 70\% 30\% &    {\scriptsize non-zero aTGCs} &   {\scriptsize  zero aTGCs} 
  &    {\scriptsize non-zero aTGCs} &  {\scriptsize  zero aTGCs}  \\ 
\hline \hline
\rule[-.6em]{0pt}{1.7em}{\scriptsize 250\,GeV(2/ab)}               & {\small $[{-4.98},{+5.14}]$}  & {\small $[{-4.68},{+4.86}]$}  &  {\small $[{-0.84},{+1.12}]$}  &  {\small $[{-0.85},{+1.11}]$}\\
\hline
\rule[-.6em]{0pt}{1.7em}{\scriptsize 250\,GeV(2/ab)+350\,GeV(200/fb)}  & {\small $[{-1.18},{+1.18}]$} & {\small $[{-1.12},{+1.12}]$}  &  {\small $[{-0.71},{+0.80}]$}   &  {\small $[{-0.69},{+0.78}]$} \\
\hline
\rule[-.6em]{0pt}{1.7em}{\scriptsize 250\,GeV(2/ab)+350\,GeV(1.5/ab)}     & {\small $[{-0.62},{+0.62}]$}   & {\small $[{-0.54},{+0.54}]$} &  {\small $[{-0.50},{+0.52}]$}  &  {\small $[{-0.47},{+0.48}]$}
\end{tabular}
\caption{One-sigma bounds on $\dkl$ from single-Higgs measurements at low-energy ILC. In this table we
consider a benchmark scenario with integrated luminosity split into $P(e^-,e^+)=(\mp 0.8, \pm 0.3)$ beam polarization with a $70\%/30\%$ ratio.}
\label{tab:resultsilcb}
\end{table}
\begin{table}[t]
\centering
\begin{tabular}{@{\hspace{4pt}}c||c|c||c|c@{\hspace{4pt}}}
$P(e^-,e^+)=(\mp 0.8, 0)$ & \multicolumn{2}{c||}{\small ILC alone} & \multicolumn{2}{|c}{\small ILC + HL-LHC}\\
\cline{2-5}
\rule[-.4em]{0pt}{1.3em} 50\% 50\% &    {\scriptsize non-zero aTGCs} &   {\scriptsize  zero aTGCs} 
  &    {\scriptsize non-zero aTGCs} &  {\scriptsize  zero aTGCs}  \\
\hline \hline
\rule[-.6em]{0pt}{1.7em}{\scriptsize 250\,GeV(2/ab)}               & {\small $[{-6.37},{+6.58}]$}  & {\small $[{-5.98},{+6.27}]$}  &  {\small $[{-0.86},{+1.13}]$}  &  {\small $[{-0.85},{+1.13}]$}\\
\hline
\rule[-.6em]{0pt}{1.7em}{\scriptsize 250\,GeV(2/ab)+350\,GeV(200/fb)}  & {\small $[{-1.40},{+1.40}]$} & {\small $[{-1.32},{+1.32}]$}  &  {\small $[{-0.74},{+0.87}]$}   &  {\small $[{-0.73},{+0.85}]$} \\
\hline
\rule[-.6em]{0pt}{1.7em}{\scriptsize 250\,GeV(2/ab)+350\,GeV(1.5/ab)}     & {\small $[{-0.71},{+0.71}]$}   & {\small $[{-0.62},{+0.62}]$} &  {\small $[{-0.55},{+0.59}]$}  &  {\small $[{-0.52},{+0.54}]$} 
\end{tabular}
\caption{One-sigma bounds on $\dkl$ from single-Higgs measurements at low-energy ILC. In this table we
consider a benchmark scenario with integrated luminosity equally split into $P(e^-,e^+)=(\mp 0.8, 0)$ beam polarization.
}
\label{tab:resultsilcc7030}
\end{table}
\begin{table}[t]
\centering
\begin{tabular}{@{\hspace{4pt}}c||c|c||c|c@{\hspace{4pt}}}
$P(e^-,e^+)=(\mp 0.8, 0)$ & \multicolumn{2}{c||}{\small ILC alone} & \multicolumn{2}{|c}{\small ILC + HL-LHC}\\
\cline{2-5}
\rule[-.4em]{0pt}{1.3em} 70\% 30\% &    {\scriptsize non-zero aTGCs} &   {\scriptsize  zero aTGCs} 
  &    {\scriptsize non-zero aTGCs} &  {\scriptsize  zero aTGCs}  \\
\hline \hline
\rule[-.6em]{0pt}{1.7em}{\scriptsize 250\,GeV(2/ab)}               & {\small $[{-5.61},{+5.83}]$}  & {\small $[{-5.27},{+5.49}]$}  &  {\small $[{-0.85},{+1.13}]$}  &  {\small $[{-0.85},{+1.13}]$}\\
\hline
\rule[-.6em]{0pt}{1.7em}{\scriptsize 250\,GeV(2/ab)+350\,GeV(200/fb)}  & {\small $[{-1.32},{+1.33}]$} & {\small $[{-1.25},{+1.25}]$}  &  {\small $[{-0.73},{+0.85}]$}   &  {\small $[{-0.72},{+0.83}]$} \\
\hline
\rule[-.6em]{0pt}{1.7em}{\scriptsize 250\,GeV(2/ab)+350\,GeV(1.5/ab)}     & {\small $[{-0.69},{+0.69}]$}   & {\small $[{-0.60},{+0.60}]$} &  {\small $[{-0.54},{+0.57}]$}  &  {\small $[{-0.50},{+0.52}]$}
\end{tabular}
\caption{One-sigma bounds on $\dkl$ from single-Higgs measurements at the low-energy ILC. In this table, we
consider a benchmark scenario with integrated luminosity split into $P(e^-,e^+)=(\mp 0.8, 0)$ beam polarization with a $70\%/30\%$ ratio.
}
\label{tab:resultsilcc5050}
\end{table}
If only ILC data are included in the fit, the precision achievable in the case of a $P(e^-,e^+)=(\mp 0.8, \pm 0.3)$ polarization with
a $70\%/30\%$ luminosity split is slightly better than the one of the other scenarios. The impact is however marginal and
basically disappears once the ILC data is combined with the high-luminosity LHC one. We find that the differences in the fits
are mainly due to the dependence of the pair production cross sections on the beam polarizations.
In \autoref{fig:hh_xsec2}, we show this dependence for the double Higgsstrahlung and $WW$-fusion pair production
cross sections. These results are obtained with \texttt{MadGraph5}~\cite{Alwall:2014hca} and do not take into account
beam-structure effects.
One can see that the largest cross sections are obtained for a $P(e^-,e^+) = (-0.8, +0.3)$ beam polarization.
The cross sections for $P(e^-,e^+) = (0, 0)$ are smaller by a factor $\sim 2$, while a much larger suppression is present for
$P(e^-,e^+) = (+0.8, -0.3)$.\footnote{Amusingly, one can note that, at leading order and independently of the center-of-mass energy, the inclusive double Higgsstrahlung production cross section with a $P(e^-, e^+) = (+0.8, -0.3)$ beam polarization configuration deviates from the unpolarized cross section by less than $1\%$.}

As a last result, we show the impact of the inclusion of the $\dkl$ parameter in the global fit on the EFT
operators. For definiteness, we focus on the circular lepton colliders benchmarks.
For the fit, we use the $12$ EFT parameters considered in the main text, namely
\begin{equation}
	\delta c_Z		\,,~~
	c_{ZZ}			\,,~~
	c_{Z\square}		\,,~~
	c_{\gamma\gamma}	\,,~~
	c_{Z\gamma}		\,,~~
	c_{gg}			\,,~~
	\delta y_t		\,,~~
	\delta y_c		\,,~~
	\delta y_b		\,,~~
	\delta y_\tau		\,,~~
	\delta y_\mu		\,,~~
	\lambda_Z		\,.
\label{eq:para12}
\end{equation}

\begin{figure}[t]
\centering
\includegraphics[width=.75\textwidth]{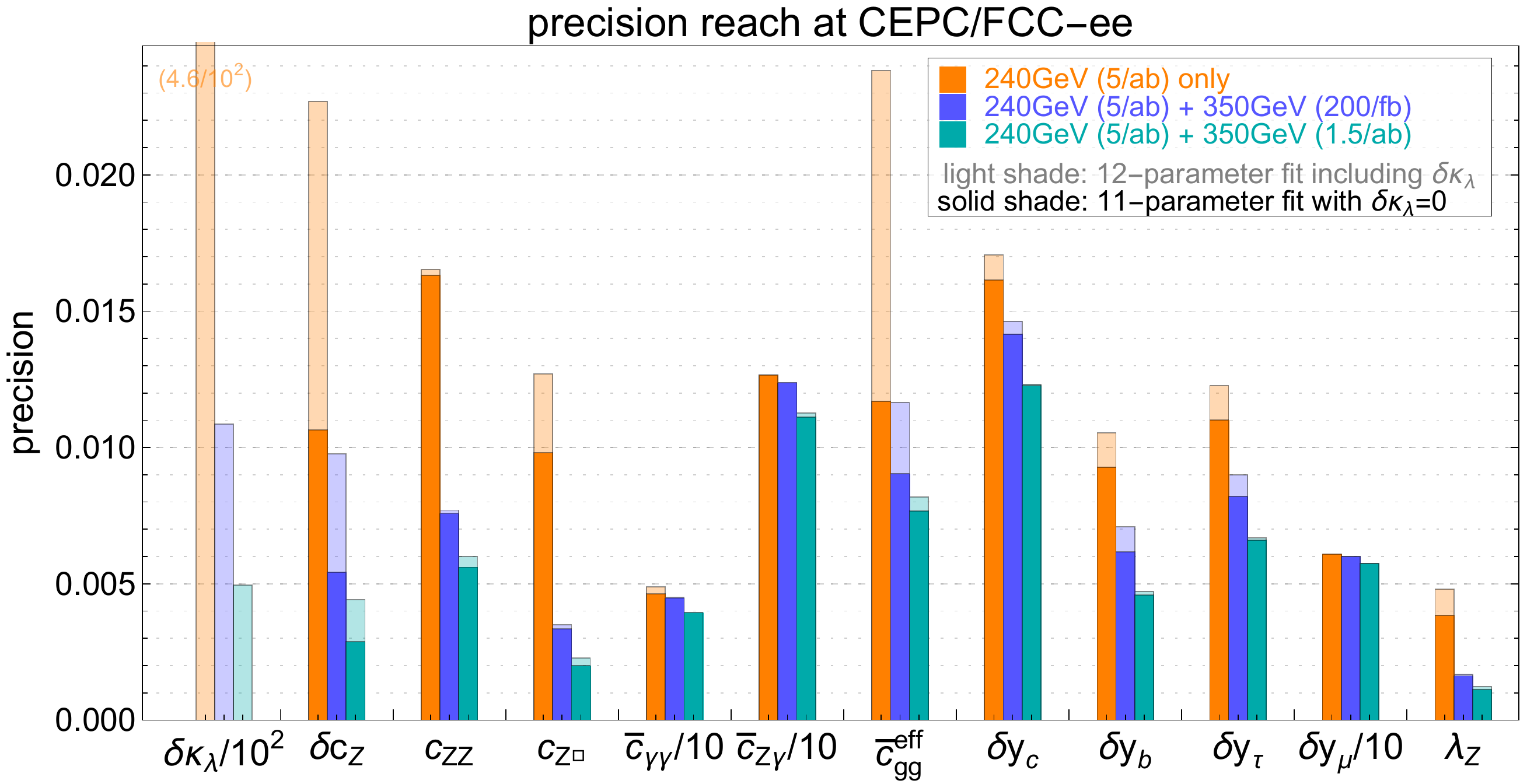}  \\ \vspace{0.5cm}
\includegraphics[width=.75\textwidth]{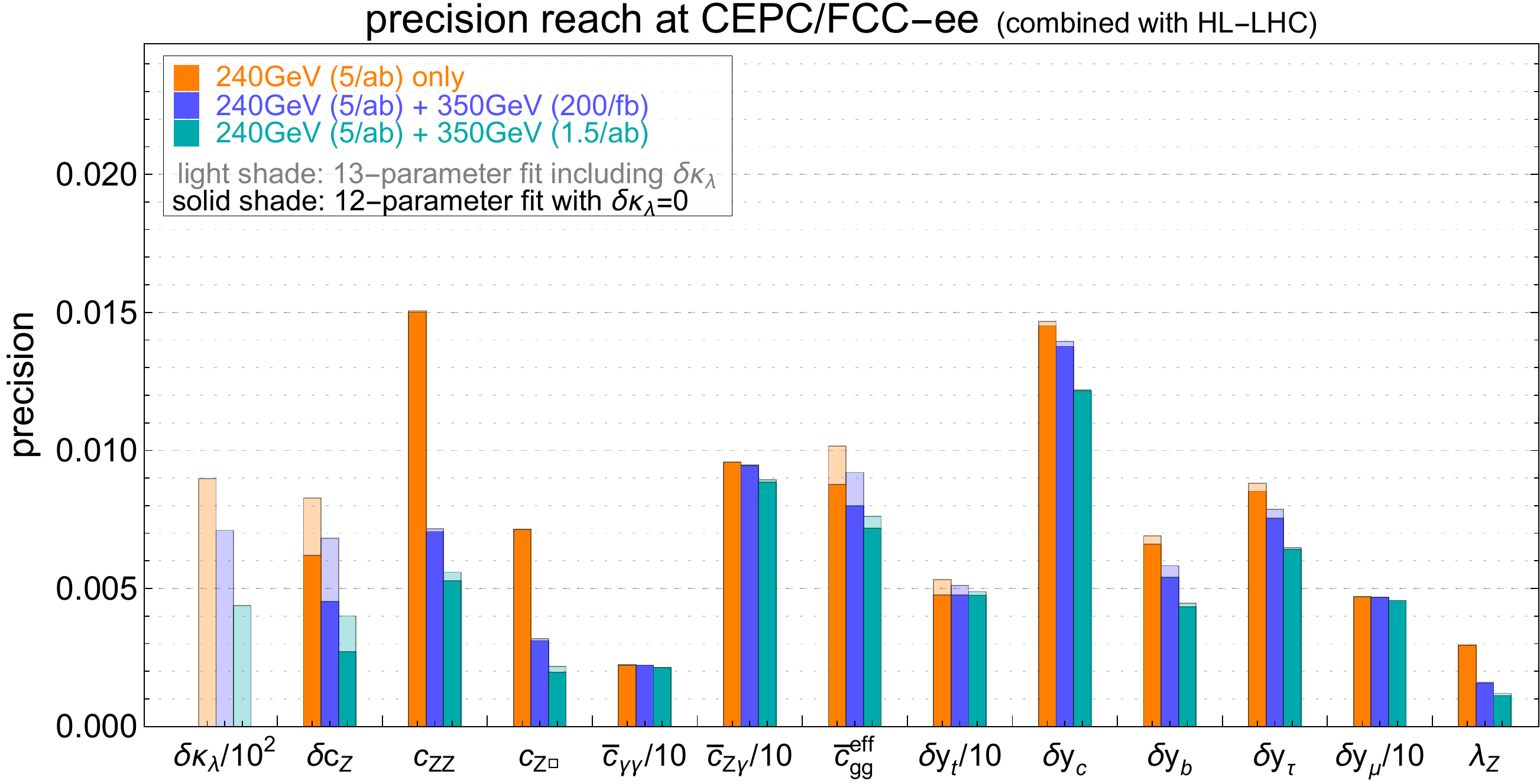}
\caption{Precision reach (one sigma constraints) at the CEPC with $5\inab$ at $240\,$GeV and $200\infb$ or $1.5\inab$ integrated luminosity at $350\,$GeV.  The upper panel shows the results of a global fit obtained from linear collider data only. The lower panel shows how the fit is modified by the inclusion of high-luminosity LHC measurements. The light-shade regions correspond to the full fit including $\dkl$, while the solid-shade regions correspond to the
fit with $\dkl = 0$.
}
\label{fig:cepclu1}
\end{figure}

As done in Ref.~\cite{Durieux:2017rsg}, it is convenient to slightly redefine the EFT parameters connected to the
Higgs decays into $\gamma\gamma$, $Z\gamma$ and $gg$. In particular we define
\begin{equation}
\frac{\Gamma_{\gamma\gamma}}{\Gamma^{\rm SM}_{\gamma\gamma}} 
	\simeq  1-2 \bar{c}_{\gamma\gamma}
	\,,  \hspace{1cm}
\frac{\Gamma_{Z\gamma}}{\Gamma^{\rm SM}_{Z\gamma}}
	\simeq 1-2 \bar{c}_{Z\gamma}
	\,,
\label{eq:barcvv}
\end{equation} 
and 
\begin{equation}
\frac{\Gamma_{gg}}{\Gamma^{\rm SM}_{gg}} 
	~\simeq~
	1 + 2\bar{c}^{\rm \,eff}_{gg}
	~\simeq~
	1+ 2 \, \bar{c}_{gg} + 2.10 \, \delta y_t -0.10 \, \delta y_b
	\,,
\label{eq:barcgg}
\end{equation}
with
\begin{equation}
\bar{c}_{\gamma\gamma}	\simeq \frac{c_{\gamma\gamma}}{8.3\times 10^{-2}} \,, \hspace{1cm} 
\bar{c}_{Z\gamma}	\simeq \frac{c_{Z\gamma}}{5.9\times 10^{-2}} \,, \hspace{1cm}  
\bar{c}_{gg}		\simeq \frac{c_{gg}}{8.3\times 10^{-3}} \,.  \label{eq:cnorm}
\end{equation}

First of all, we focus on the fit obtained from low-energy lepton colliders only. In this case, the top Yukawa coupling
and the Higgs contact interaction with gluons can not be accessed independently and can only be tested through
the Higgs decay into $gg$. The $\delta y_t$ and $c_{gg}$ parameters thus always appear in combination as shown
in \autoref{eq:barcgg}. In the global fit we include only the $\bar c_{gg}^{\rm eff}$ parameter and not $c_{gg}$ and $\delta y_t$
separately. The precision on the various EFT parameters with and without the inclusion of $\dkl$ is shown
in the upper panel of \autoref{fig:cepclu1}. One can see that, if only a $240\,$GeV run is available, the inclusion
of the Higgs self-coupling in the fit significantly degrades the precision on $\delta c_Z$ and $\bar c_{gg}^{\rm eff}$.
In this case, as we already discussed in the text, the precision on $\dkl$ is very low. The situation changes drastically
in the presence of runs at $350\,$GeV. In this case, the precision on $\bar c_{gg}^{\rm eff}$ is effectively decoupled from
the determination of the Higgs trilinear coupling. Some correlation of $\dkl$ with $\delta c_Z$ is still present
with $200\infb$ of integrated luminosity at $350\,$GeV, while a much milder effect remains with $1.5\inab$ of integrated luminosity.

In the lower panel of \autoref{fig:cepclu1}, we show the global fit obtained after combination with high-luminosity LHC measurements. In this case, the top Yukawa and the Higgs contact interaction with gluons can be independently tested.
The results of the global fit show that the inclusion of the Higgs trilinear coupling affects only the determination
of $\delta c_Z$. The impact is however much smaller than in the fit with lepton collider data only. Other EFT parameters are affected in a negligible way.

\providecommand{\href}[2]{#2}\begingroup\raggedright\endgroup

\end{document}